\documentclass[a4paper,11pt,final,manyauthors]{fundam}
\usepackage{cite}

\usepackage{tabularx}
\usepackage{amssymb,amsfonts,amscd,amsmath}
\usepackage{xspace}
\usepackage[misc]{ifsym}
\usepackage{xkvltxp}
\usepackage{fixme}
\usepackage[pdftex]{graphicx}

\newcommand{\andb}{\mbox{\bf AND}\xspace}
\newcommand{\orb}{\mbox{\bf OR}\xspace}
\newcommand{\sandb}{\mbox{\bf SAND}\xspace}
\newcommand{\semant}[1]{\mbox{$[\![#1]\!]$}}

\newcommand{\semParM}[1]{\semant{#1}_M}
\newcommand{\atSet}{\mbox{$\cal AT$}\xspace}
\newcommand{\clTok}{\mbox{\sf Tok}}
\newcommand{\clTyp}{\mbox{\sf Typ}}
\newcommand{\DL}{\mbox{${\mathcal FD}$}\xspace}
\newcommand{\symbolw}[1]{\mbox{\sf #1}}
\newcommand{\EoP}{\hfill$\Box$\vspace*{\baselineskip}}
\newcommand{\Ifmtag}{IM}

\newtheorem{Prop}{{\bf Proposition}}
\newtheorem{Def}{{\bf Definition}}
\newtheorem{Lem}{{\bf Lemma}}
\newtheorem{Cor}{{\bf Corollary}}

\title{On Validating Attack Trees with Attack Effects: An Approach
  from Barwise-Seligman's Channel Theory\thanks{This paper is based on
    the proceeding presented in SAFECOMP 2020~\cite{safecomp20} (see
    Section~\ref{Intro_Result} in detail).}}
\author{Hideaki Nishihara\\
  SEI-AIST CyberSecurity Cooperative Research Lab.\\
  National Institute of Advanced
  Industrial Science and Technology (AIST), Osaka, Japan\\
  h.nishihara@aist.go.jp\\~
  \and
  Yasuyuki Kawanishi\thanks{
    Also affiliated at SEI-AIST CyberSecurity Cooperative Research Lab.
    in AIST.}\\
  Cyber-Security R\&D Office\\ Sumitomo Electric Industries, Ltd.,
  Osaka, Japan\\
  kawanishi-yasuyuki@sei.co.jp
  \and
  Daisuke Souma\thanksas{2}
  \\Cyber-Security R\&D Office\\ Sumitomo Electric Industries, Ltd.,
  Osaka, Japan\\
  souma-daisuke@sei.co.jp
  \and
  Hirotaka Yoshida\\
  SEI-AIST CyberSecurity Cooperative Research Lab.\\
  National Institute of Advanced
  Industrial Science and Technology (AIST), Osaka, Japan\\
  hirotaka.yoshida@aist.go.jp
}
\address{SEI-AIST CyberSecurity Cooperative Research Lab.,
  National Institute of Advanced
  Industrial Science and Technology (AIST), Osaka, Japan.}
\runninghead{H. Nishihara et al.}{On Validating Attack Trees with Attack Effects}

\begin{document}
\maketitle
\begin{abstract}
  In security analysis,
  attack trees are a major tool for showing the structural
  decomposition of attacks and for supporting the evaluation of the
  quantitative properties (called attributes) of the attacks.
  However, the validities of decompositions are not established by
  attack trees themselves, and fallacious decisions about security may
  be made when the attack trees are inaccurate.  This paper enriches
  attack trees with effects of attacks, with a formal system focusing
  on refinement scenarios.  Relationships among effects indicate
  relationships among attacks and it allows for a systematic
  evaluation of attack decompositions.  To describe effects this paper
  applies Barwise-Seligman's channel theory.  Infomorphisms, in
  particular, play a significant role to connect effects with distinct
  granularities.
  As a result, the consistency of a decomposition is formally defined
  and a condition for it is stated.
  This framework is applied to a case study of a vehicular network
  system.  As an application of the idea of consistency, possible
  degrees of mitigation for attacks in attack trees are discussed.
\end{abstract}



\section{Introduction}
\subsection{Background}\label{Background}
Progress in information technology has led to the evolution of various
systems worldwide.  In particular, cyber-physical systems now have
more flexible and finer functionalities, and cooperate with other
systems via networks.  However, security threats to these systems have
also increased, and protecting against them has become an important
issue recently.

Attack trees are a major tool in analyzing the security of a
system~\cite{Schneier,EVITA09,ATMedical,GuidedDesign}, as they
represent the decomposition of threats in the form of
\andb/\orb-trees, alongside fault trees that represent the structures
of faults in a system in the safety domain.  As shown in the following
example and Section~\ref{ATSect}, attack trees match formal approaches
in that they enable us to understand a threat's logical structures and
estimate its qualitative/quantitative properties.

\begin{figure}[hbtp]
  \begin{center}
      \includegraphics[keepaspectratio,width=80mm]{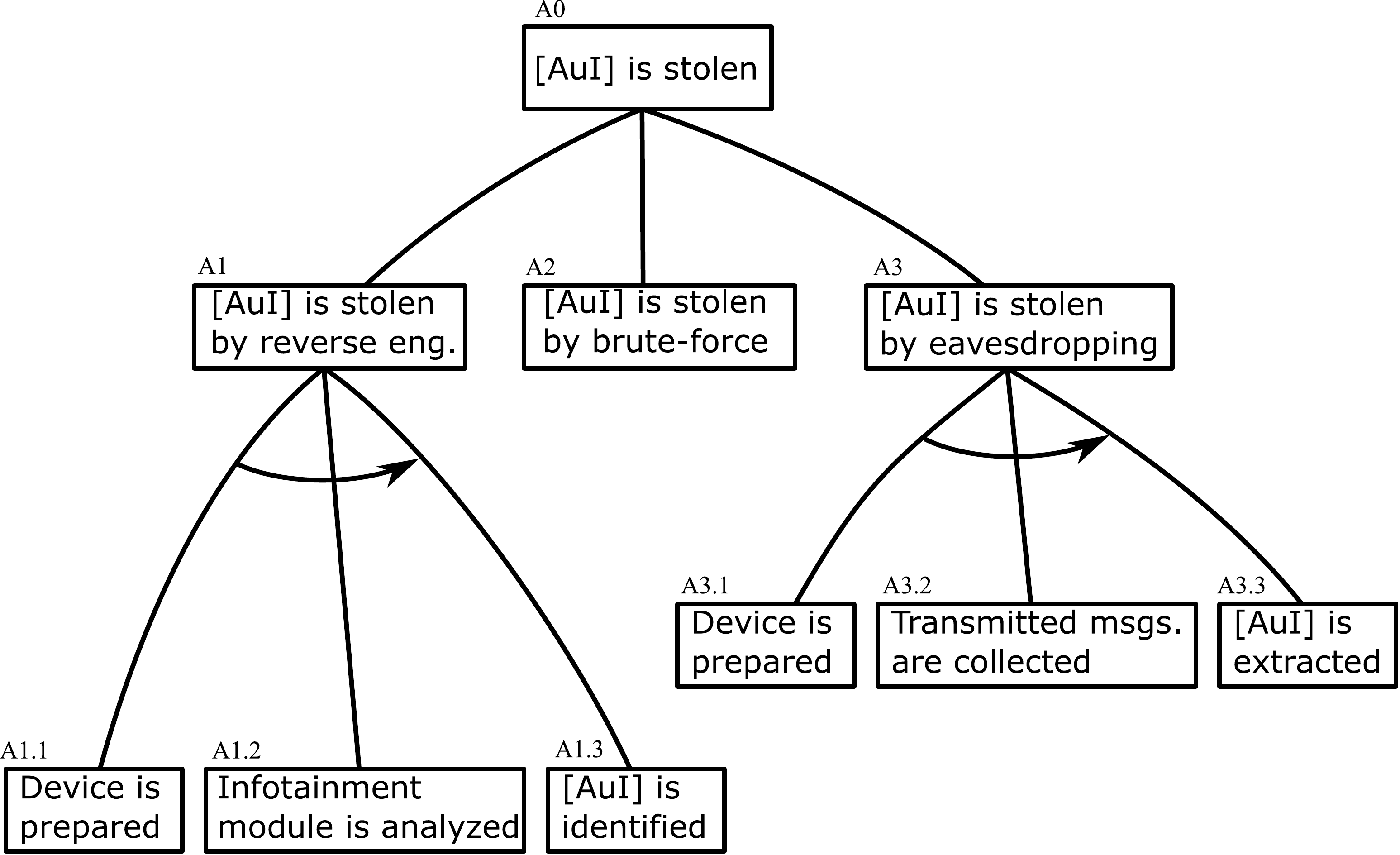}
  \end{center}
  \caption{Attack tree}\label{ATeg0}
\end{figure}

Fig.~\ref{ATeg0} is an example of an attack tree representing the
threat ``authentication information in Infotainment module ([AuI] for
short) is stolen'' to a vehicle systems (see also
Section~\ref{ToESubSect}).  If the threat is realized, then a
malicious user can connect their smartphone to Infotainment and
control it or access critical subsystems in the vehicle.  In the
diagram, the threat to be analyzed is placed at the top node (the
root) of the tree, and is decomposed into sub-attacks recursively.
The branch just below the root node is an \orb branch.  When at least
one child of an \orb branch is done successfully, the threat or attack
of the parent node is considered to be done.  The branch below node A1
is a \sandb branch.  In order for the attack of the parent node of the
branch to be considered successful, all of the child nodes must be
done in order from left to right.  We draw a small arrow around a
\sandb branch.  Further, we can use \andb branches in attack trees.
As with \sandb branches, all of the child nodes of an \andb branch are
required for the attack of the parent node to be successful, but the
order of their execution does not matter.

Fig.~\ref{ATeg0} shows three attack scenarios to achieve the threat on
the top: reverse engineering of Infotainment module(A1), brute-force
attack(A2), and eavesdropping(A3).  Scenario A1 is concretized to the
sequence of three actions: preparing the device that contains
Infotainment module(A1.1), analysis of the module (A1.2), and
identifying [AuI](A1.3).  Scenario A3 is concretized to the action
sequence as well.  When some attributes are assigned to each node at
the bottom (the leaf), we can inductively estimate the attributes of
the upper nodes, especially the root node.  For example, let us
consider the possibility of an attack.  If the actions labeled A1.1,
A1.2, A1.3, A3.1, and A3.3 are possible, then the intermediate action
A1 is possible naturally, while actions A2 and A3 are not..  This
implies the action of the root A1 is possible as the result.

Simple and intuitive descriptions of attack trees allow various
extensions of the concepts (see also Section~\ref{ATov}).  Examples
include adding other types of nodes, connecting trees expressing
defenses, and specifying the maximum number of children of a node.

Fortunately, attack trees are well formulated in
research~\cite{MO,SeqConjHMT,SeqConjJKMRT,ATcorrect}; formal syntax
and semantics are provided, the quantitative attributes are related to
attacks formally, or attacks are linked to state transitions of the
target system.

\subsection{Problem and Result}\label{Intro_Result}
Not all aspects of attack trees are supported by previous
formulations.  In particular, the relations among attacks around a
branch have not been discussed thoroughly.  In the literature on
attack trees, one branch represents a concretization of the parent,
another branch shows the preceding events for the parent, and these
ideas are mingled together at the other branch.  The resulting attack
trees tend to be diverse, which poses a problem from a practical
viewpoint.  Although several studies~\cite{ATcorrect,GuidedDesign,CID}
have addressed to this issue, the frameworks they proposed are rather
indirect.  Therefore, methodologies need to be developed for
considering the consistency of attack trees simply.

This paper tackles this problem by introducing the effects of attacks.
An effect, considered a post-condition of an event, is tightly related
to an attack, and therefore we can use the concept of effects to
discuss the relationships among attacks around a branch.  Furthermore,
when one attack depends on another one, they are related with effects;
that is, the effect of the preceding attack functions as a
pre-condition of the subsequent attack.  Accordingly, the
decomposition of an attack must correctly depict the relationships
among effects, and equivalently relationship violations among effects
reveal an inadequate decomposition of the attack.  Establishing this
idea is the primary goal of this paper.

We apply the theory of information flow and channel, developed by
Barwise and Seligman~\cite{BSInfoFlow}, and define the consistency of
a branch in terms of effects.  The theory is originally used to
describe whole-part relations of concepts appearing in distributed
systems.  A {\it classification} is assigned to each component in the
system and they are linked with specific mappings called {\it
  infomorphisms} to the classification for the entire system.  It
means the components and their properties of the system with distinct
granularities are dealt with in a united way.  By the theory, the
effects of nodes in an attack tree, that inhabit in some
classifications, are ``lifted'' into the classification of their
parent node.  If the lifted effects imply the parent's effect, then
the latter is regarded as the abstraction of the child's effect.  On
one hand, {\it integrations} of plural effects, that can reside in
distinct classifications, are defined.  Therefore, around a branch in
an attack tree we can compare the integration of child nodes' effects
and the parent's effect to check their abstraction-refinement
relation.  It leads to the definition of the consistency of the branch
and this approach allows us assess the validity of attack
decompositions in formal way.  Moreover, it supports rigorous analyses
of attack trees, including evaluations of the completeness of the
decompositions.

We define a novel semantics of attack trees in order to consider
intermediate nodes.  Horne et al.~\cite{SeqConjHMT} provided a
semantics of attack trees with sequential conjunctions, which takes
values in a set of directed graphs whose nodes are labeled with
primitive attacks.  Here, a primitive attack corresponds to a leaf
node in an attack tree.  It means that attack trees are interpreted as
combinations of only primitive attacks.  However, this paper focuses
on the relationships among an attack and its sub-attacks, especially
at intermediate nodes.  We consider that an attack tree expresses a
collection of inseparable refinement scenarios.  The semantics
proposed in this paper takes value in the powerset of sub-trees
without \orb branches.  These sub-trees can derive directed graphs
labeled by the leaf nodes in the attack tree, and therefore the
semantics can be related to the semantics proposed by Horne et al.
(summarized in Appendix~\ref{SectProjCAT}).

As potential applications of attack trees with effects, we evaluate
countermeasures and possible mitigation for attacks.  A countermeasure
or mitigation eliminates some of the consequences (i.e., effects) of
attacks.  Hence, the mitigated effects and the residuals can be
described as fragments of effects.  This enables us to link the
mitigation to the consistency of the attack decomposition.  We
formalize the idea, and check that the residual effects of the
sub-attacks tend to be stronger than the residual effect of the
parent.  Specifically, by looking at a case study of a vehicular
network system and potential threats to it, we analyzes possible
countermeasures in detail from this viewpoint.

Generally, an attack tree process for decomposing a threat consists of
three steps: identification of the target, tree construction, and
analysis.  The first step is commonly carried out in system
engineering, such as system development or risk management.  To
conduct this step in a systematic way, we can follow established
methodologies such as with SysML~\cite{SysML}.  On the one hand, for
the third step, formal analyses with attack trees have been developed.
With the use of attributes, logical or quantitative properties of a
threat are integrated according to the tree structure, enabling a
rigorous assessment of the threat(see also Section~\ref{ATov}).
However, the second step of the process, namely, a systematic approach
to tree construction, seems to be overlooked, as we pointed out
ambiguous relationships among attacks.  Our results support tree
constructions by observing consistency in a more direct, simpler and
formal way.  As a result, all steps in the attack tree process can be
approached systematically, helping to improve the attack tree
analysis.

This paper is an extension of~\cite{safecomp20}.  Theoretical supports
for the ideas and results constitute the main addition
(Section~\ref{EffTr}).  Relationships between attacks and effects are
described based on the concept model and are formalized with
Information flow theory.  These justify the discussion on the
refinement of attacks such as with abstract-refinement relations of
effects, integrations of effects, and mitigation.  Explicitly, the
major differences are as follows:
\begin{itemize}
\item An introductory example is added (Section~\ref{Background});
\item Symbols that express the types of branches are simplified in diagrams;
\item The formal definition is compared with~\cite{SeqConjHMT} in detail
  (Section~\ref{ATdef} and Appendix~\ref{SectProjCAT});
\item A concept model of effects is built, and the discussion treating
  effects is formalized with Information flow theory
  (Section~\ref{SectConceptModel});
\item The definition of consistency (Definition~\ref{DefConsistency})
  is generalized, with special consideration of the residual effects
  around \sandb branches;
\item The validity of decompositions is mentioned explicitly
  (Corollary~\ref{ATValidity});
\item The discussion on mitigation is separated to a theoretical part
  (Section~\ref{SectMitigation}) and a case study (Section~\ref{EvalSubSect});
\item The discussion on treatment levels for effects is removed since it
  does not fit into Information flow framework;
\item Examples in the case study are re-described with the updated
  theoretical framework.
\end{itemize}

\paragraph*{Organization.}
Section~\ref{ATSect} explains the general theory of attack trees and
includes an overview of related works.  It also provides a formal
definition of attack trees with sequential conjunctions and
attributes.  Section~\ref{EffTr}, the main part of this paper,
introduces the concept of effects of attacks and discusses the
consistency of branches in attack trees.  First, the requirements for
effects are identified.  Next, a framework to describe effects is
established based on Information flow theory.  Then, the consistency
of branches is defined formally, together with integrations of child
nodes' effects.  Possible mitigations of attacks are also discussed in
this section.  Based on this discussion, Section~\ref{CStudy}
illustrates a case study on threats to a vehicular network system.
The examples show how a decomposition of attacks is verified, and
possible mitigation of attacks.  Section~\ref{Conclusion} concludes
the paper.  A detailed explanation of how the formal system of attack
trees in this paper derives the system by Horne et al. is shown in
Appendix~\ref{SectProjCAT}.

\section{Attack Trees with Sequential Conjunctions}\label{ATSect}
\subsection{Overview of Attack Trees}\label{ATov}
In this section, we review attack trees, particularly the formal
descriptions as well as their practical applications.  Comparisons
with fault trees used in the safety domain are also discussed.

The concept of attack trees was firstly introduced by
Schneier~\cite{Schneier}, who expressed the decomposition of an attack
as an \andb/\orb-tree and demonstrated several examples of evaluating
of attacks using the tree.  That is, he evaluated an attack by
integrating the evaluations of sub-attacks along with the tree
structure.  Subsequently, this idea was formalized by Mauw and
Oostdijk~\cite{MO}, who specified a formal syntax of attack trees and
defined the corresponding semantics as a set of multisets consisting
of primitive attacks.  Moreover, they discussed the equivalence and
transformations of attack trees compatible with the semantics and
called the evaluations of attacks as {\it attributes}.  An attribute
was defined as a function from the nodes of an attack tree to a set,
where the function values did not conflict with \andb and \orb
decomposition.  Examples of attributes are possibility of
attacks and the costs for attacks~\cite{Schneier}, attack time and
the minimum number of experts for attacks~\cite{SeqConjHMT}, and
attack probabilities~\cite{EVITA09}.
The attribute of an attack in an attack tree
was calculated with the attribute values of the children.

Attack trees are applied in various domains for the purpose of attack
modeling, although in many cases, they are not defined formally.
Recently, security analyses with attack trees were conducted for
cyber-physical systems.  One study~\cite{ATMedical} showed an attack
tree for an implantable medical device and investigated whether
communication protocols for the device had vulnerabilities or not.
Another study~\cite{UrbanRail} analyzed the security for a railway
system.  Attack trees were applied to identify detailed attack
scenarios, while effect identification and risk evaluation was done
with Failure Modes, Vulnerabilities, and Effects Analysis (FMVEA).  In
EVITA~\cite{EVITA09} for the automotive domain, attack trees were
considered as major tools to identify attack scenarios and to estimate
attack potentials.  However, approaches to building trees were not
discussed apart from abstract tree structures.  JASO TP
15002~\cite{TP15002} for automobiles suggested tree decompositions of
selected threats to analyze how the threats could be realized.  In
DO-356~\cite{DO-356} for the aviation domain, tree diagrams were
introduced to analyze security aspects.  They were called threat
trees, as they were focused on threat condition events and
vulnerability events as well as attacks.

The idea of attack trees is rather simple, and allows various
extensions.  Wang et al.~\cite{UParamAT} classified many variants of
attack trees, and Fovino and Masera~\cite{FM} enriched attack tree
nodes with related information such as assertions, vulnerabilities,
and operations.  With such enrichment, attacks or threats can be
analyzed from several viewpoints.  The simplicity of attack trees also
allows wider interpretations, which means practitioners may experience
difficulties in building them.  To the best of our knowledge, most
related studies have explicitly discussed neither guidance for attack
decompositions nor the validity of decompositions in detail.  Although
the research\cite{ATcorrect,GuidedDesign,CID} discussed this issue,
their frameworks dealt with attacks only indirectly.

One of the major extensions of attack trees was the addition of a new
branch type, namely, sequential conjunction.  In several cases,
sub-attacks of an attack have causal dependency, and therefore it is
natural to consider an attack tree together with the order of attack
executions.  Attack trees with sequential conjunction were discussed
by Jhawar et al.~\cite{SeqConjJKMRT} and Horne et
al.~\cite{SeqConjHMT}.  Their studies extended Mauw and Oostdijk's
formalization~\cite{MO}, in particular, the semantics was extended
from multisets to sets of graphs representing possible sequences of
primitive attacks.  Audinot et al.~\cite{ATcorrect} also focused on
attack trees with sequential conjunctions.  They used pre- and
post-conditions of attacks for the labels of nodes instead of the
attacks themselves.  The semantics was given as the sets of the
target's behaviors that satisfied the corresponding pre- and
post-conditions.  The framework clearly shows the relationships among
the parent node and its children around a branch, but actual events by
attacks are not presented explicitly.  Furthermore, a transition model
expressing possible behaviors in the target system should be prepared
in advance.  Audinot et al. discussed the consistency\footnote{It is
called {\it correctness} properties in their paper.} of decomposition
by comparing the semantics of nodes around a branch.  Andr{\'{e}} et
al.~\cite{ParamAFT} formulated attack trees as representations of
event sequences about attacks.  A timed automaton is assigned to each
leaf in an attack tree and the automata of lower nodes are composed at
the upper nodes.  Around a branch, the parent node activates the child
nodes depending on the branch type, executes the automata assigned to
the activated children in parallel or in series, and finally
integrates their results.

In the safety domain, fault trees have been used for the reliability
analysis of systems since the 1960s.  Fault trees were presented as
\andb/\orb-trees the same as attack trees, but they showed causal
decompositions~\cite{IEC61025, RUIJTERS}.  The literature often
treated attack trees and fault trees similarly.  For example,
interpretations were given as sets of labels on the leaf nodes in the
tree, or properties of the uppermost event of the tree were
quantitatively calculated with the properties of leaf
nodes~\cite{MO,RUIJTERS}.  Moreover, with a recent observation that
security threats caused harms in safety-critical systems, the
integration of attack trees and fault trees was proposed~\cite{FM2} to
connect security analysis and safety analysis.

\subsection{Formulation}\label{ATdef}
We provide a formal definition of attack trees.  The syntax is defined
inductively, and the semantics represents possible scenarios of attack
refinements.  In the sequel, we denote attack trees, including
sequential conjunctions.

\begin{Def}\label{ATSyntax}
An attack tree is a labeled tree with three types of branches:
\begin{eqnarray*}
  t&::=&Lf(n)~|~Nd(n,op,\langle t,t,\dots,t\rangle),\\
  op&::=&\andb~|~\orb~|~\sandb,
\end{eqnarray*}
where $\langle -\rangle$ means a non-empty finite sequence of its
arguments and the symbol $n$ is a label for the node, which expresses
an action or event.  The set of attack trees is denoted by \atSet.
\end{Def}

\begin{figure}[hbtp]
  \begin{center}
      \includegraphics[keepaspectratio,width=100mm]{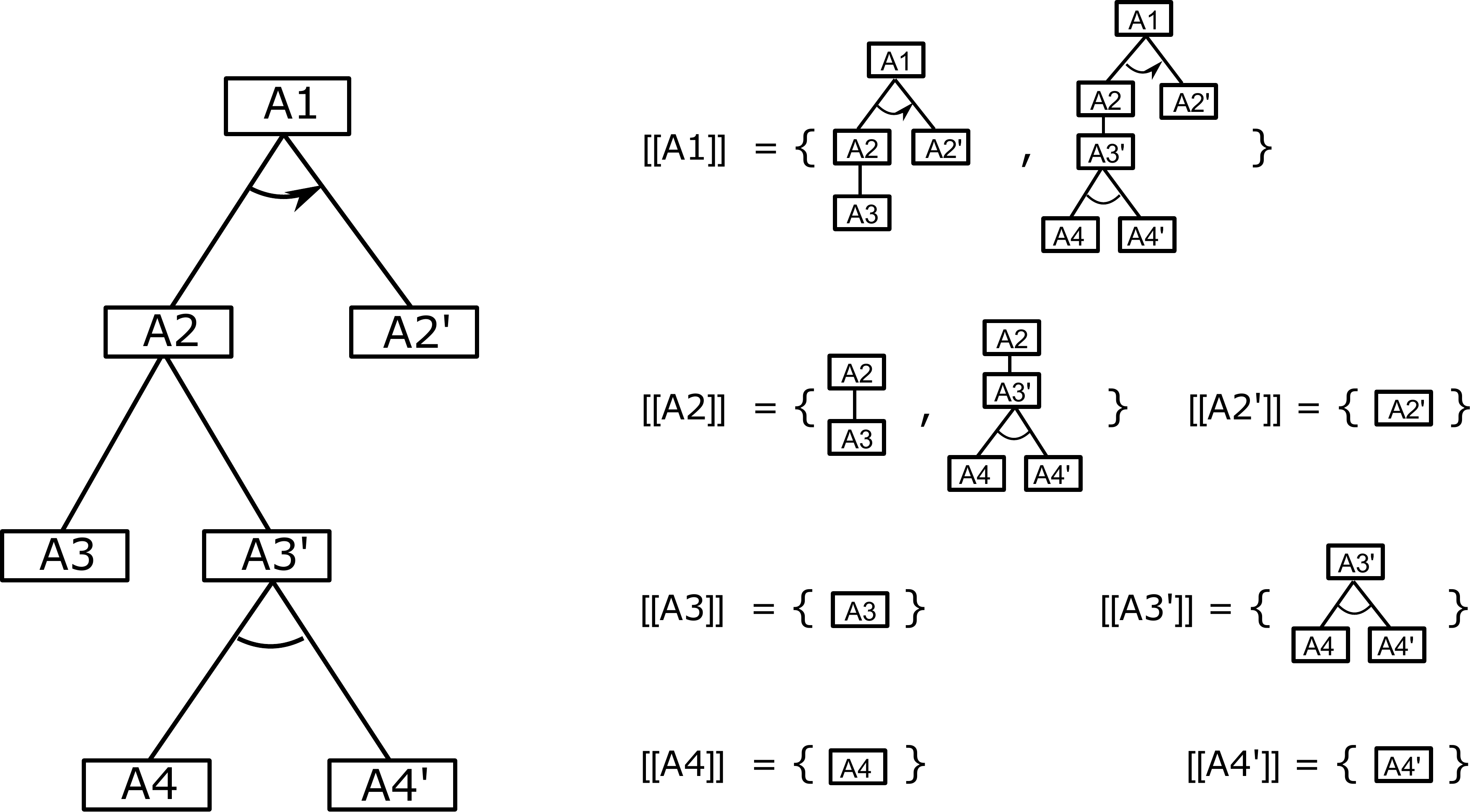}
  \end{center}
  \caption{Attack tree and its interpretation}\label{ATeg}
\end{figure}

Intuitively, $Lf(n)$ corresponds to a primitive attack $n$, which is
no longer decomposed (a {\it leaf} node in a tree), and
$Nd(n^\prime,op,\langle t_1,\dots,t_k\rangle)$ corresponds to an
attack $n^\prime$, which has sub-trees $t_1,\dots,t_k$ with type $op$
as its decomposition (an intermediate node in a tree).  Attack trees
can be diagrammatically represented, as illustrated in
Fig.~\ref{ATeg}.  The branch type is expressed around it: an arc for
\andb branch, a small arrow for \sandb branch, and no symbols for
\orb branch.  The child nodes around a \sandb branch are executed from
left to right.  The uppermost node of an attack tree is called the
root node.

We do not consider the order of children for \andb or \orb branches,
and we do not care the branch type when it has only one child.  Hence
the following equalities are assumed for arbitrary subtrees
$\{t_j\}_{1\le j\le k}$ and a subtree $t$:
\begin{eqnarray*}
  \lefteqn{Nd(n,op,\langle t_1,\dots,t_i,t_{i+1},\dots,t_k\rangle)}\\
  &~~~~=&Nd(n,op,\langle t_1,\dots,t_{i+1},t_i,\dots,t_k\rangle)
  ~~(op\in\{\andb,\orb\})\\
  \lefteqn{Nd(n,op,\langle t\rangle)=Nd(n,op^\prime,\langle t\rangle)
  ~~(op,op^\prime\in \{\andb,\sandb,\orb\})}.
\end{eqnarray*}

We denote an attack tree without \orb branches as an {\it
  R-tree}\footnote{ It corresponds to a {\it codot term}
in~\cite{SeqConjHMT}.}.  Intuitively, an R-tree expresses an
individual refinement scenario concerning the attack of the root node.
The set of R-trees is denoted by $\atSet_R$.

A semantics $\semant{\cdot}$ of attack trees is the function that
maps an attack tree to a multiset of R-trees.
\begin{Def}\label{ATSemantics}
  The function $\semant{\cdot}$ on \atSet is
  defined by the following rules, where $\bar{t}=\langle t_1,\dots,t_m\rangle$
  and $\semant{\,\bar{t}\,}=(\semant{t_1},\dots,\semant{t_m})$:
\begin{eqnarray*}
\semant{Lf(n)} &=&\{Lf(n)\},\\
\semant{Nd(n,\andb,\bar{t} )}
  &=&\{Nd(n,\andb,\langle \tau_1,\dots,\tau_m\rangle)~|~
  (\tau_1,\dots,\tau_m)\in\semant{\,\bar{t}\,}\},\\
\semant{Nd(n,\sandb,\bar{t})}
&=&\{Nd(n,\sandb,\langle \tau_1,\dots,\tau_m\rangle)~|~
  (\tau_1,\dots,\tau_m)\in\semant{\,\bar{t}\,}\},\\
\semant{Nd(n,\orb,\bar{t})}&=&
\bigsqcup_{1\le i\le m}\{Nd(n,\andb,\langle\tau\rangle)|\tau\in\semant{t_i}\}.
\end{eqnarray*}
\end{Def}

The semantics is based on the idea that decompositions in
attack trees are logical refinement.  An \orb branch is interpreted as
a multiset union, indicating the branch corresponds to a case
division.  An aspect of the attack is refined, and possible detailed
attacks are listed as sub-attacks.  On the other hand an \andb/\sandb
branch is interpreted as a factorization of an attack to sub-attacks.
The collection of sub-attacks around the branch is inseparable, as a
single sub-attack in it does not invoke the original attack.  The causal
dependency of attacks exists only between the children of each \sandb
branch and does not exist elsewhere, especially between an attack and its
sub-attacks.

\paragraph{Remark.}
A comparison of our formulation of attack trees (Definition~\ref{ATSyntax},
\ref{ATSemantics}) with those of {\it causal attack trees}
in~\cite{SeqConjHMT} (reviewed in Appendix~\ref{HMTDef}) shows the
syntaxes are very similar -- the difference is whether or not a
branch is limited to having two children.  On one hand, our semantics keeps
non-leaf nodes, and analyses of them are available, whereas the
``intermediate semantics'' for causal attack trees only considers leaf
nodes.  Therefore, we can {\it project} attack trees to causal ones,
as stated in Proposition~\ref{PropProjCAT} in Appendix~\ref{ProofProjCAT}.
As a result, the attack trees in \atSet can be analyzed by
structural methods for causal attack trees, after their consistencies
are confirmed as discussed in Section~\ref{EffTr}.

\subsection{Attributes}
An attribute of an attack tree is defined as a function $f$ from the
set of nodes.  The codomain of $f$ depends on the node in general, but
most of the known attributes have fixed common codomains like numbers
or boolean values.  Each codomain of $f$ is a set with the three
operations $\mu_O$, $\mu_A$, and $\mu_S$, corresponding to \orb,
\andb, and \sandb, respectively.  Around a branch in an attack tree,
attribute values of the child nodes are summarized with these
operations.  Therefore, the following equalities are required to hold:
\begin{eqnarray*}
\lefteqn{\mu_O(\langle f(x_1),\dots,f(x_i),f(x_{i+1}),\dots,f(x_k)\rangle)}\\
&~~~~&=\mu_O(\langle f(x_1),\dots,f(x_{i+1}),f(x_i),\dots,f(x_k)\rangle),\\
\lefteqn{\mu_A(\langle f(x_1),\dots,f(x_i),f(x_{i+1}),\dots,f(x_k)\rangle)}\\
&~~~~&=\mu_A(\langle f(x_1),\dots,f(x_{i+1}),f(x_i),\dots,f(x_k)\rangle),\\
\lefteqn{\mu_A(\langle f(x)\rangle)=\mu_S(\langle f(x)\rangle)
  =\mu_O(\langle f(x)\rangle).}
\end{eqnarray*}

One example of attributes is the minimum number of experts required to
perform an attack, as discussed in~\cite{SeqConjHMT}.  When we denote
the defining function of the attribute by $\nu$, its codomain is
defined as the set of natural numbers $\mathbb{N}$ and
$(\mu_O,\mu_A,\mu_S)=(\min,\mbox{Sum},\max)$.  Note that this
attribute is assumed to be determined by the values of lower nodes.
Namely, $\nu(Nd(n,\orb,\bar{t})) =\min\{\nu(t_1),\dots,\nu(t_k)\}$
where $\bar{t}=\langle t_1,\dots,t_k\rangle$, and similar equations
hold for \andb/ \sandb branches.  When there is another attribute for
which we cannot make this assumption, the equalities may not be
expected and we must consider contributions for the attribute by the
intermediate nodes themselves.  Such an attribute will be compatible
with the semantics in Definition~\ref{ATSemantics} but not with that
in~\cite{SeqConjHMT} using only leaf nodes.  To distinguish the latter
type of attribute, we often call them {\it quasi-attributes}.
We see an example of a quasi-attribute in Section~\ref{ConstBranch}.

\section{Validating Decompositions with Effects}\label{EffTr}
\subsection{Effects of Attacks}\label{EffOfAttack}
An effect is one of the major properties of an attack.  It is a
situation or property of a specific entity related to the target
system and is caused by a specific action.  For example, consider the
attack ``{\it Message receive function is interfered}.''  After the
attack, messages may be lost, the function may be unavailable, or
other irregular behaviors may occur.  As these situations did not
occur before the attack, it can be considered that the attack caused
them.  As a result, we can identify the summarized situation ``{\it
  Messages are not processed correctly}'' as the effect of the attack.

Effects are also significant concepts in the areas adjacent to
security.  In ISO/IEC Guide 51~\cite{Guide51:2014} safety standard,
the primary issues to avoid are negative effects\footnote{Those
effects are referred to as harms.}  on people, property, or the
environment caused by some events.  In ISO 31000~\cite{iso31000:2018}
standard focused on risk management, a risk is defined as an effect of
uncertainty on objectives.

Owing to these observations,
it seems reasonable that {\it an effect of an attack} meets
the following conditions:
\begin{itemize}
\item The effect must be directly caused by the corresponding attack.
  It shows a change of the entity that the attack affects;
\item Properties that hold before the attack must not be selected as
  effects;
\item The effect occurs immediately after the corresponding attack,
  and no other properties invoked by the attack occur before the
  effect occurs.
\end{itemize}

\begin{figure}[htbp]
  \begin{center}
    \includegraphics[width=70mm]{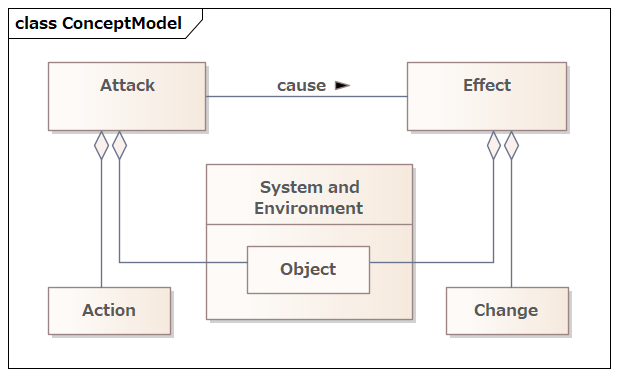}
    \caption{Concept model}\label{ConceptModel}
  \end{center}
\end{figure}

Fig.~\ref{ConceptModel} is a model of the relationship between an attack
and an effect.  An attack consists of the target object
and action for the target, while an effect is a change of the
target.  The object is related to the target system or its
environment, and the target of an attack coincides to the object of an
effect.

The effects of attacks can be described by logical formulas.  For
example, the effect ``{\it The password becomes public or is
  overwritten}'' has the target {\it password}, and its changes are being public
and being overwritten.  Hence, the effect is intuitively
decomposed to the pair of symbols \symbolw{passwd} and
$\symbolw{Disclosed}\lor\symbolw{Modified}$.

\subsection{Refinement and Information Flow}\label{SectConceptModel}

In order to discuss effects and their refinement according to the
model in Fig.~\ref{ConceptModel}, we apply Information flow and
Channel theory~\cite{BSInfoFlow} and describe several concepts and
relationships.

Intuitively, a system consists of several subsystems and they progress
in cooperation with each other.  Each subsystem has its individual
state at an arbitrary time and some properties of it hold depending on
the state.  The states and properties of the entire system are
represented as integrations of states and properties of subsystems.
This idea is formalized by the theory of information flows and
channel.  Classifications defined in Channel theory describe
relations between properties and situations on a system and its
subsystems, while infomorphisms in the theory interlink elements in
the classifications.  With an infomorphism, a property of a
subsystem is ``transferred'' into another subsystem and 
analyzed together with properties in the second subsystem.

Channel theory is used in development of cyber-physical systems.
Gebreyohannes~\cite{FormalReqChannel} applied Channel theory to model
communications among a satellite and ground stations.  The
communication possibility to the satellite, which changes by time and
distance, were caught by classifications and infomorphisms.
Peters~\cite{Peters04} presented a hierarchical model of a swarmbot.
Components in the model with different granularities were linked
with infomorphisms.

We start with reviewing basic definitions.

\begin{Def}
  A classification $C$ is a triple $(\clTok(C),\clTyp(C),\vDash)$
  where $\clTok(C)$ (`tokens' of $C$) and $\clTyp(C)$ (`types' of $C$) are
  sets, and $\vDash$ is a binary relation between them.
\end{Def}

\begin{Def}
  Let $C_1$ and $C_2$ be classifications.  An infomorphism from $C_1$
  to $C_2$ is a pair of mappings $(f^\wedge,f^\vee)$, where
  $f^\wedge:\clTyp(C_1)\to\clTyp(C_2)$,
  $f^\vee:\clTok(C_2)\to\clTok(C_1)$, and the following condition is
  satisfied:
  \[
  f^\vee(a)\vDash_1\gamma\iff a\vDash_2 f^\wedge(\gamma)\tag{\Ifmtag}
  \]
  for all $a\in\clTok(C_2)$ and
  $\gamma\in\clTyp(C_1)$.  We write this infomorphism as
  $(f^\wedge,f^\vee):C_1\rightleftarrows C_2$.
\end{Def}

\paragraph{Example.}
Consider a classification $W_1$ and suppose
\begin{eqnarray*}
  \clTok(W_1)&=&\{\symbolw{passwd},\symbolw{pTimeout}\},\\
  \clTyp(W_1)&=&\{\symbolw{Disclosed}, \symbolw{Modified},\symbolw{Hidden}\},
  and\\
  \symbolw{passwd}\vDash\symbolw{Disclosed},&~&
  \symbolw{pTimeout}\vDash\symbolw{Modified}
\end{eqnarray*}
that formalizes the status of the password and the timeout parameter
in a software module.  In addition, consider another classification
$W_2$ and suppose
\begin{eqnarray*}
  \clTok(W_2)&=&\{\symbolw{auth}\},\\
  \symbolw{pass-through}&\in&\clTyp(W_2),\\
  \symbolw{auth}&\vDash&\symbolw{pass-through}, \mbox{\rm and}\\
  \symbolw{auth}&\not\vDash&\alpha (\mbox{\rm if $\alpha\in\clTyp(W_2)$
    is not \symbolw{pass-through}})
\end{eqnarray*}
that formalizes the status of the authentication function
(the lowest relation says ``\mbox{\it Authentication is not functioning}'').
Now take mappings $f^\wedge:\clTyp(W_1)\to\clTyp(W_2)$ and
$f^\vee:\clTok(W_2)\to\clTok(W_1)$ with
\begin{eqnarray*}
  f^\wedge(\symbolw{Disclosed})&=&\symbolw{pass-through},\\
  f^\wedge(\symbolw{Modified})&\neq&\symbolw{pass-through}\neq
  f^\wedge(\symbolw{Hidden}), \mbox{\rm and}\\
  f^\vee(\symbolw{auth})&=&\symbolw{passwd}.
  \end{eqnarray*}
Then the pair $(f^\wedge,f^\vee)$ is an infomorphism.\EoP

Infomorphisms represent {\it whole-part} relationships.  Now let us
see the classification $W_2$ modeling a function of a software system,
and $W_1$ modeling its components.  The relation
$\symbolw{passwd}\vDash_1\symbolw{Disclosed}$ in $W_1$, whose token
comes from the concepts of the entire function (i.e., it has a
preimage by $f^\vee$), is {\it lifted} to the relation
$\symbolw{auth}\vDash_2\symbolw{pass-through}$ in $W_2$ with the
infomorphism $(f^\wedge,f^\vee)$.  We later apply this idea to
representing abstract-refinement relations of effects.

Compositions of effects are required for our purpose; however, the
original framework of classifications is not suitable for that,
especially since related pairs of tokens and types cannot be composed
when tokens are distinct.  Now we develop some compound
classifications.  To begin, we suppose a set of indices $\Omega$,
where any index set appearing in this paper is its subset.

\begin{Def}\label{CompoundCla}
  Let $C$, $C_1$, and $C_2$ be classifications that have finite tokens
  and types.  They also have `un-connected tokens' $\varepsilon$,
  $\varepsilon_1$, $\varepsilon_2$ respectively.  Namely,
  $\varepsilon\not\vDash\alpha$ for $\forall \alpha\in\clTyp(C)$ and
  similar conditions hold for $\varepsilon_1$ and $\varepsilon_2$.
  \begin{enumerate}
  \item The classification $C_1\oplus C_2$ is defined as
    \begin{itemize}
    \item $\clTok(C_1\oplus C_2)=\clTok(C_1)\sqcup\clTok(C_2)$
      (disjoint union)
    \item $\clTyp(C_1\oplus C_2)=\clTyp(C_1)\sqcup\clTyp(C_2)$,
    \item $a\vDash_{C_1\oplus C_2}\alpha$ holds if and only if either of
      the following statements holds
      \begin{itemize}
      \item $a^\prime\vDash_1\alpha^\prime$ for some $a^\prime\in\clTok(C_1)$
        and $\alpha^\prime\in\clTyp(C_1)$ such that $a=in_1(a^\prime)$ and
        $\alpha=in_1(\alpha^\prime)$, where $in_1$ is the embedding to
        the disjoint union, or
      \item $a^{\prime\prime}\vDash_2\alpha^{\prime\prime}$ for some
        $a^{\prime\prime}\in\clTok(C_2)$ and $\alpha^{\prime\prime}\in\clTyp(C_2)$
        such that $a=in_2(a^{\prime\prime})$ and $\alpha=in_2(\alpha^{\prime\prime})$,
        where $in_2$ is the embedding to the disjoint union.
      \end{itemize}
    \end{itemize}
  \item The classification $(C_1,C_2)$ is defined as\footnote{
  The assumption of $\varepsilon_i$ can be ignored to construct $(C_1,C_2)$.
  }
    \begin{itemize}
    \item $\clTok((C_1,C_2))=\clTok(C_1)\times\clTok(C_2)$,
    \item $\clTyp((C_1,C_2))=\clTyp(C_1)\times\clTyp(C_2)$,
    \item $\langle a_1,a_2\rangle\vDash_{(C_1,C_2)}\langle \alpha,\beta\rangle
      \iff$ $a_1\vDash_1\alpha$ and $a_2\vDash_2\beta$.
    \end{itemize}
  \item The classification $\DL(C)$ is about `family tokens' and `distributive
    lattice types' generated by $C$.  It is defined as follows.
    \begin{itemize}
    \item A token of $\DL(C)$ is a family $\{a_\lambda\}_{\lambda\in\Lambda}$
      where $a_\lambda\in\clTok(C)$ and $\Lambda\subset\Omega$ is finite.
      In the sequel, we often write tokens like as $\{a_\lambda\}_\Lambda$
      when it does not invoke some confusion.
    \item A primitive type of $\DL(C)$ is presented as $\alpha_\lambda$ where
      $\alpha\in\clTyp(C)$ and $\lambda\in\Omega$.
      A generic type is constructed as
      \[
      \tau::=\chi~ |~ \top~ |~ \bot~
      |~\tau\land\tau ~|~ \tau\lor\tau 
      ~(\mbox{$\chi$ is a primitive type}).
      \]
      $\clTyp(\DL(C))$ is subject to relations\footnote{ Relations are
      required to be compatible with token-type relation:
      $\forall\{a_\lambda\}_\Lambda.\{a_\lambda\}_\Lambda\vDash\rho\iff
      \{a_\lambda\}_\Lambda\vDash\sigma$ if $\rho$ and $\sigma$ are related.}
      of distributive lattices and semantical
      relations among primitive types, that is the resulting
      lattice is not always a free-generated\footnote{Precisely, the
      structure of $\clTyp(\DL(C))$ depends on the relations and is
      not unique.  However we fix one structure for every $C$ in this
      paper, and thus we denote the lattice by the abused name
      $\clTyp(\DL(C))$.}.
    \item For a primitive type $\alpha_\mu$,
      $\{a_\lambda\}_{\lambda\in\Lambda}\vDash_{\DL(C)}\alpha_\mu \iff
      \mu\in\Lambda\mbox{~and~}a_\mu\vDash_C\alpha$.  Moreover
      $\{a_\lambda\}_\Lambda\vDash\top$ and
      $\{a_\lambda\}_\Lambda\not\vDash\bot$ for arbitrary token
      $\{a_\lambda\}_\Lambda$.
    \item For compound types,
      \[
      \{a_\lambda\}_\Lambda\vDash_{\DL(C)}\Gamma\land\Delta\iff
      \{a_\lambda\}_\Lambda\vDash_{\DL(C)}\Gamma \mbox{ ~and~}
      \{a_\lambda\}_\Lambda\vDash_{\DL(C)}\Delta.
      \]
      Similarly, $\{a_\lambda\}_\Lambda\vDash_{\DL(C)}\Gamma\lor\Delta$ is defined.
    \end{itemize}
  \end{enumerate}
\end{Def}

Two types of relations are introduced in $\DL(C)$-type classifications.
First, there are several structural deductions of relations concerning tokens:
\begin{itemize}
\item $\{a_\lambda\}_{\Lambda}\vDash\gamma_\mu\iff
  \{a_\lambda\}_{\Lambda-\{\mu^\prime\}}\vDash\gamma_\mu$ if $a_\mu=a_{\mu^\prime}$
  where $\mu,\mu^\prime\in\Lambda$.
\item $\{a_\lambda\}_{\Lambda}\vDash\Gamma\iff
  \{a_\lambda\}_{\Lambda-\{\rho\}}\vDash\Gamma$ if $\rho$ has no occurrence
  in any primitive type in $\Gamma$.
\item $\{a_\lambda\}_{\Lambda}\vDash\Gamma\iff
  \{a_\lambda\}_{\Lambda-\{\rho\}}\vDash\Gamma$ if $a_\rho=\varepsilon$.
\end{itemize}

Second, a distributive lattice structure defines a partial orders on
$\clTyp(\DL(C))$, i.e.,
$\Gamma\le\Delta\iff\Gamma\lor\Delta=\Delta\iff
\Gamma\land\Delta=\Gamma$.  We interpret this order as a derivation
such that ``the greater type can be derived from the smaller type.''
As an example, consider that $\Gamma\land\Delta\le\Gamma$ means
$\Gamma$ can be derived from $\Gamma\land\Delta$.  Hence, we often
write $\Theta\Rightarrow Z$ instead of $\Theta\le Z$.  Moreover,
phenomena in the modeled world are reflected as relations in the
classification.  For example,
$(\symbolw{passwd}\vDash\symbolw{Disclosed})
\Rightarrow(\symbolw{passwd}\vDash\symbolw{Accesseible})$ is assumed
since the disclosure of the password includes the situation that the
password is accessible.  Introducing relations like this leads us to
take a quotient of the distributive lattice $\clTyp(\DL(C))$.

We review general properties of the classifications defined above.
First we see the construction of $\DL(C)$ is so-called functorial.
Let $(f^\wedge,f^\vee):
C_1\rightleftarrows C_2$ be an infomorphism.  We can define
\begin{eqnarray*}
  \DL
  f^\wedge:\clTyp(\DL(C_1))&\to&\clTyp(\DL(C_2)):\\ \alpha_\lambda&\mapsto&
  f^\wedge(\alpha)_\lambda,\\ \Gamma\sharp\Delta&\mapsto& \DL
  f^\wedge(\Gamma)\sharp\DL f^\wedge(\Delta)
\end{eqnarray*}
where $\sharp\in\{\land,\lor\}$.  We assume that this map is compatible with
relations among primitive types and is well-defined.  On one hand,
we can define
\begin{eqnarray*}
\DL f^\vee:\clTok(\DL(C_2))\to\clTok(\DL(C_1))&:&
\{a_\lambda\}_\Lambda\mapsto\{a^\prime_\lambda\}_\Lambda
\end{eqnarray*}
where $a^\prime_\lambda=f^\vee(a_\lambda)$.

\begin{Lem}\label{FBfunctorial}
  Let $(f^\wedge,f^\vee):C_1\rightleftarrows C_2$ and
  $(g^\wedge,g^\vee):C_2\rightleftarrows C_3$ be infomorphisms between
  classifications $C_1$, $C_2$, and $C_3$.
  \begin{enumerate}
  \item The construction above defines an infomorphism
    $(\DL f^\wedge,\DL f^\vee):\DL(C_1)\rightleftarrows\DL(C_2)$.
  \item $\DL(g^\wedge\circ f^\wedge)=\DL g^\wedge\circ\DL f^\wedge$
    and $\DL(g^\vee\circ f^\vee)=\DL g^\vee\circ\DL f^\vee$ hold.  As
    well, $\DL(id_C^\wedge)=id^\wedge_{\DL(C)}$ and
    $\DL(id_C^\vee)=id^\vee_{\DL(C)}$ hold.
  \item  $\DL(f^\wedge)$ and $\DL(g^\wedge)$ are order-preserving.
  \end{enumerate}
\end{Lem}

Statements 2 and 3 are obvious, and Statement 1 is derived from the
following equivalence:
\begin{eqnarray*}
  \lefteqn{\DL f^\vee(\{a_\lambda\}_\Lambda)\vDash\gamma_\mu
    ~~\mbox{(in $\DL(C_1)$)}}\\
  &\iff&\mu\in\Lambda\mbox{~and~}f^\vee(a_\mu)\vDash\gamma
  ~\mbox{in $C_1$}\\
  &\iff&\mu\in\Lambda\mbox{~and~}a_\mu\vDash f^\wedge(\gamma)
  ~\mbox{in $C_2$}\\
  &\iff&\{a_\lambda\}_\Lambda\vDash_{\DL(C_1)}\DL f^\wedge(\gamma_\mu)
  ~~\mbox{(in $\DL(C_2)$)}.
\end{eqnarray*}

Here remark that, for $\DL(C)$-type classifications, it is enough to check
the condition of infomorphisms (\Ifmtag) for generators of $\clTyp(C)$ only.

As stated in the next lemma, $\DL(C)$ and $C_1\oplus C_2$ are related
to the original classifications $C$, $C_1$, and $C_2$, with
infomorphisms\footnote{On one hand, relating $C_i$ and $(C_1,C_2)$
requires additional conditions.  However, since the relationship
between them does not appear in this paper, we do not consider the
matter any further.}. We write $in_i(x)$ as $x^{(i)}$ for short.

\begin{Lem}
  \begin{enumerate}
  \item Fix an index  $\mu\in\Omega$.
    The $\mu$-th embedding and the $\mu$-th projection
    \begin{eqnarray*}
      \mbox{\rm lift}^\wedge_\mu:\clTyp(C)\to\clTyp(\DL(C))&:
      &\alpha\mapsto\alpha_\mu,~\mbox{and}\\ \mbox{\rm
        lift}^\vee_\mu:\clTok(\DL(C))\to\clTok(C)&:&
      \{a_\lambda\}_{\Lambda}\mapsto\left\{\begin{array}{ll} a_\mu &
      (\mu\in\Lambda)\\ \varepsilon & (\mbox{\rm otherwise})
    \end{array}\right.
    \end{eqnarray*}      
    constitute an infomorphism $C\rightleftarrows\DL(C)$.    
  \item For $i\in\{1,2\}$, the pair of the functions
    \begin{eqnarray*}
      \mbox{\rm inc}^{i\wedge}(\zeta)&=& in_i(\zeta)=\zeta^{(i)},
      ~\mbox{and}\\
      \mbox{\rm inc}^{i\vee}(x)&=&\left\{\begin{array}{ll}x^\prime &
      (\mbox{\rm if there exists $x^\prime$ s.t. $x=in_i(x^\prime)$})\\
      \varepsilon & (\mbox{\rm otherwise})\end{array}\right..
    \end{eqnarray*}
    is an infomorphism $C_i\rightleftarrows C_1\oplus C_2$.
  \end{enumerate}
\end{Lem}
It is not difficult to check whether functions in the lemma satisfy
the condition (\Ifmtag).

As a corollary of Lemma~\ref{FBfunctorial}, the next proposition
describes $\DL(C_1\oplus C_2)$.  Notice that a token of $\DL(C_i\oplus
C_2)$ is expressed as a family $\{a^{(k_\lambda)}_\lambda\}_{\Lambda}$.
The parameter $k_\lambda$ indicates the component where the token
$a_\lambda^{(k_\lambda)}$ belongs in the original; that is, $a_\lambda\in\clTok(C_i)$
if $k_\lambda=i$.
\begin{Prop}\label{PropLift}
  The pair $(\overline{\mbox{\rm inc}^i}^\wedge,\overline{\mbox{\rm inc}^i}^\vee)
  =(\DL\ \mbox{\rm inc}^{i\wedge},\DL\ \mbox{\rm inc}^{i\vee})$ is
  an infomorphism $\DL(C_i)\rightleftarrows\DL(C_1\oplus C_2)$.
  For $i\in\{1,2\}$,
  \begin{eqnarray*}
    \overline{\mbox{\rm inc}^i}^\wedge(\alpha_\lambda)&=&(\alpha^{(i)})_\lambda,
    ~\mbox{and}\\
    \overline{\mbox{\rm inc}^i}^\vee(\{a_\lambda^{(k_\lambda)}\}_{\Lambda})&=&
    \{a_\lambda\}_{\Lambda_i}
  \end{eqnarray*}
  for $\Lambda_i=\{\lambda\in\Lambda|k_\lambda=i\}$.
  Especially,
  $\overline{\mbox{\rm inc}^i}^\vee(\{a_\lambda^{(k_\lambda)}\}_{\Lambda})$ is the empty
  family if $\Lambda_i=\emptyset$.
\end{Prop}

The next proposition says that the classification $(\DL(C_1),\DL(C_2))$
is embedded into $\DL(C_1\oplus C_2)$.

\begin{Prop}\label{PropLift2}
  The mappings on $(\clTyp(\DL(C_1)),\clTyp(\DL(C_2)))$ and
  $\clTok(\DL(C_1)\oplus\DL(C_2))$ defined as
  \begin{eqnarray*}
  \overline{\mbox{\rm conj}}^\wedge(\langle\Gamma,\Delta\rangle)&=&
  \Gamma^{(1)}\land\Delta, \mbox{and}\\
  \overline{\mbox{\rm conj}}^\vee(\{a^{(k_\lambda)}_\lambda\}_{\Lambda})
  &=&\langle\{a_\lambda\}_{\Lambda_1},\{a_\lambda\}_{\Lambda_2}\rangle
  \end{eqnarray*}
  constitute an infomorphism $(\DL(C_1),\DL(C_2))
  \rightleftarrows\DL(C_1\oplus C_2)$, where
  $\Gamma^{(1)}$ [re. $\Delta^{(2)}$] is obtained by replacing all
  primitive type like $\alpha$ with $\alpha^{(1)}$ [re. $\alpha^{(2)}$], and
  where $\Lambda_i=\{\lambda\in\Lambda|k_\lambda=i\}$.
  Moreover this infomorphism is mono.  i.e.
  if $\overline{\mbox{\rm conj}}\circ g=
  \overline{\mbox{\rm conj}}\circ h$ then $g=h$ holds where
  $\overline{\mbox{\rm conj}}=
  (\overline{\mbox{\rm conj}}^\wedge,\overline{\mbox{\rm conj}}^\vee)$.
\end{Prop}

\paragraph{Example.}
Remember the classifications $W_1$ and $W_2$ in the last example.  Additionally
we introduce a new classification $W_3$:
\begin{eqnarray*}
  \symbolw{dest}&\in&\clTok(W_3)\\
  \symbolw{Modified}&\in&\clTyp(W_3)\\
  \symbolw{dest}&\vDash&\symbolw{Modified}
\end{eqnarray*}
In $W_1$, (or $W_2$, $W_3$ as well) we can only describe relationships
between single tokens and single types.  Now relationships for plural
tokens and types, especially transversal ones, can be described in
single relations in $\DL(\oplus W_i)$.  Concretely, under
$a_1=\symbolw{passwd}$, and $a_2=\symbolw{pTimeout}$,
\[
\{a_1,a_2\}\vDash\symbolw{Disclosed}_1\land
\symbolw{Modified}_2.
\]
is a relation held in $\DL(W_1)$.  Moreover, this relation is lifted to
(with $a_3=\symbolw{dest}\in\clTok(W_3)$)
\[
\{a_1^{(1)},a_2^{(1)},a_3^{(3)}\}
\vDash\symbolw{Disclosed}^{(1)}_1\land\symbolw{Modified}^{(1)}_2
\]
by $(\overline{\mbox{\rm inc}^i}^\wedge,\overline{\mbox{\rm inc}^i}^\vee)$.
Like that, the relation $\symbolw{dest}\vDash_3\symbolw{Modified}$
is lifted to
\[
\{a_1^{(1)},a_2^{(1)},a_3^{(3)}\}\vDash\symbolw{Modified}^{(3)}_3,
\]
and finally they are composed to the relation in $\DL(W_1\oplus W_3)$:
\[
\{a_1^{(1)},a_2^{(1)},a_3^{(3)}\}
\vDash(\symbolw{Disclosed}^{(1)}_1\land\symbolw{Modified}^{(1)}_3)
\land\symbolw{Modified}^{(3)}_2.
\]
\EoP

Now we formalize the refinement of relations in classifications.  Let
$\DL(C_C)$ and $\DL(C_A)$ be classifications and $D$ be another
classification embedded in $\DL(C_C)$.  Consider an infomorphism
$(f^\wedge,f^\vee):D\rightleftarrows\DL(C_A)$.  When a relation
$(\vec{a}\vDash\Gamma)$ in\footnote{ We often write a token in a
$\DL(C)$-type classifications as a vector to emphasize that it is a
family of $C$ 's tokens.}  $\DL(C_C)$ belongs to the embedded image of
$D$ and the token $\vec{a}$ has a preimage by $f^\vee$, it can be
lifted to $(\vec{a}^\prime\vDash_A f^\wedge(\Gamma))$ via the
embedding, and we can compare it with other relations in $\DL(C_A)$.
\begin{Def}
  In this situation, we say
  $(\vec{a}^\prime\vDash_A\Delta)$ is an abstraction of
  $(\vec{a}\vDash_C\Gamma)$ (or $(\vec{a}\vDash\Gamma)$ is a refinement of
  $(\vec{a}^\prime\vDash\Delta)$) by $(f^\wedge,f^\vee)$,
  if the following implication holds:
  \[
  (\vec{a}^\prime\vDash_Af^\wedge(\Gamma))\Rightarrow(\vec{a}^\prime\vDash_A\Delta)
  \]
  i.e., $f^\wedge(\Gamma)\le\Delta$ on the token $\vec{a}^\prime$.
\end{Def}

\subsection{Consistent Branches}\label{ConstBranch}
As shown in Fig.~\ref{ConceptModel}, we consider objects in the target
system and its environment, and a specific attack affects some of the
objects.  Remark that these objects depend on the granularity of the
attack; tampering with a Telematics module does not target the
authentication subfunction directly but rather the entire
communication services.

For an attack $A$, let us consider a classification $C_A$.
$\clTok(C_A)$ is the set of objects in the target system and its
environment such that their granularities are in keeping with the
attack $A$.  $\clTyp(C_A)$ is the set of properties about some
elements in $\clTok(C_A)$.  For $a\in\clTok(C_A)$ and
$\gamma\in\clTyp(C_A)$, $a\vDash\gamma$ holds if and only if $\gamma$
is an effect on $a$, i.e., $\gamma$ is a property of $a$ satisfying
the three conditions addressed in Section~\ref{EffOfAttack}.  With
$C_A$ and the semantical relations between its types, we can consider
the classification $\DL(C_A)$ and compound properties of the attack
$A$.  The order of $\clTyp(\DL(C_A))$ expresses the strength of
effects.  In our context, effects mean some negative impacts, and
thus, holding plural properties indicates stronger effects.  For
example, $(\vec{a}\vDash\Gamma_1)$, $(\vec{a}\vDash\Gamma_2)$, and
$\Gamma_1\le \Gamma_2$ mean $\Gamma_1=\Gamma_1\land\Gamma_2$ and thus
$\Gamma_1$ is stronger than $\Gamma_2$.

\begin{figure}
  \begin{center}
    \includegraphics[keepaspectratio,width=\textwidth]{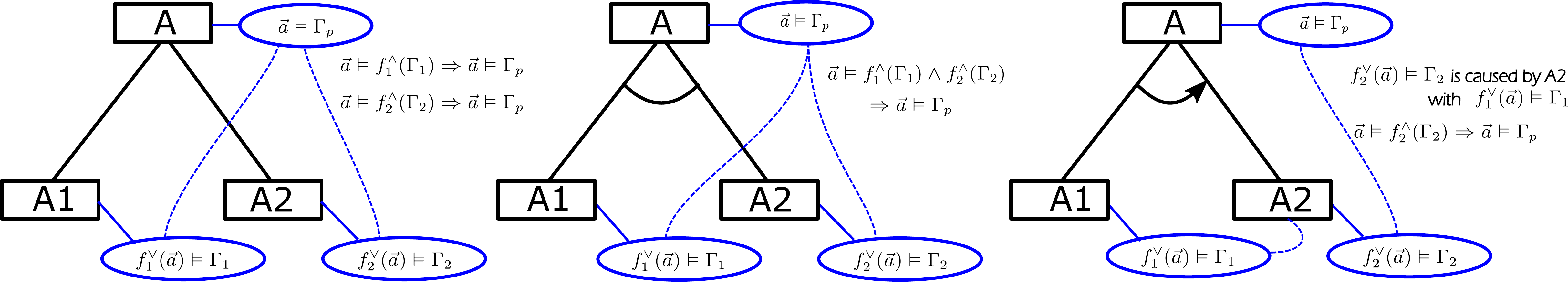}
  \end{center}
  \caption{Consistency of attacks with effects}\label{ATconsistency}
\end{figure}

Now we can assign an effect to each node of an attack tree.  For a
node\footnote{We abuse nodes and their labels, unless any confusion is
not invoked.} $N$ in the tree, choose a pair $(\vec{a},\Gamma)\in
\clTok(\DL(C_N))\times\clTyp(\DL(C_N))$ such that
$\vec{a}\vDash\Gamma$ holds, and regard it as the effect of $N$.  In
diagrams, we put the round node labeled by the effect around $N$ and
connect them with a blue edge (see Fig.~\ref{ATconsistency}).

Since a branch in an adequately constructed attack tree represents a
refinement of the attack on the parent node, a similar structure can
be expected for effects due to the concept model (Fig.~\ref{ConceptModel}).
For instance, the effect of the
parent node will be derived from the conjunction of all effects of the child
nodes for an \andb branch, because all of the attacks corresponding to the
child nodes are executed.  By contrast, when there are several
conflicts among the effects around a branch, the decomposition of the
attack will have inconsistencies, such as a misunderstanding of the
situation or inadequate refinement.

To describe refinement around \sandb branches, we introduce {\it cut
  sequences}.  For a sequence
$\langle\vec{a}^1\vDash\Gamma_1,\vec{a}^2 \vDash\Gamma_2,
\dots,\vec{a}^n\vDash\Gamma_n\rangle$, we pick up the rightmost
elements with respect to each token and form a new sub-sequence
$\langle\vec{a}^{i_1}\vDash\Gamma_{i_1},\vec{a}^{i_2}\vDash
\Gamma_{i_2},\dots,\vec{a}^{i_k}\vDash\Gamma_{i_k}\rangle$
($i_1<i_2<\dots<i_k$) as the cut sequence.  Namely, if tokens
$\vec{a}^i$ and $\vec{a}^j$ ($i<j$) in the original sequence are
equivalent, then the element $\vec{a}^i\vDash\Gamma_i$ is removed from
the sequence.  By definition, all tokens in a cut sequence are
distinct.

The consistencies of branches in attack trees are defined by the use of
infomorphisms.  Consider a branch in an attack tree and
denote by $p$ [re. $i$] an index for the
parent node [re. the $i$-th child node].  For instance, $C_p$ and
$(\vec{a}^p\vDash\Gamma_p)$ express the classification and the effect
assigned to the parent node, respectively.
\begin{Def}\label{DefConsistency}
  A branch with $n$ child nodes in an attack tree is called
  consistent, if the condition below holds regarding its type:
    \begin{itemize}
    \item \orb branch: For each index $i$, there exists an infomorphism
      $f_i:D_i\rightleftarrows\DL(C_p)$ such that
      the $i$-th effect is a refinement of the parent node's effect
      by $f_i$, where $D_i$ is an embedded classification in $\DL(C_i)$
      to which the $i$-th effect belongs.
    \item \andb branch: There exists an infomorphism
      $f:D\rightleftarrows\DL(C_p)$ such that the tuple of child's
      effects is a refinement of the parent node's effect by $f$,
      where $D$ is an embedded classification in $(\DL(C_i))_{1\le i\le n}$
      to which the tuple of child's effects belongs.
    \item \sandb branch: The two conditions below hold.
      \begin{itemize}
      \item Every $i$-th effect
        $(\{a^i_\lambda\}_{\lambda\in\Lambda_i}\vDash\Gamma_i)$ is
        obtained by the $i$-th attack with assuming the effects in the cut
        sequence of the preceding effects
        $\langle\{a^1_\lambda\}_{\lambda\in\Lambda_1}\vDash\Gamma_1$, $\dots$,
        $\{a^{i-1}_\lambda\}_{\lambda\in\Lambda_{i-1}}\vDash\Gamma_{i-1}\rangle$.
      \item For the cut sequence of the tuple of child's effects,
        there exists an infomorphism $f$ from an embedded classification
        $D$ in $(\DL(C_r))_{r\in I}$ to $\DL(C_p)$ such that the same
        conditions for \andb branches hold, where $I$ is the index set
        of the cut sequence.
      \end{itemize}
    \end{itemize}
    These conditions are depicted in Fig.~\ref{ATconsistency}.

    If all branches in an attack tree are consistent, then the
    entire attack tree is called consistent.
\end{Def}
Examples of attack trees with effects are provided in Section~\ref{CStudy}.

Let us denote the mapping from nodes in an attack tree to their
effects by $\varphi$.  The codomain of $\varphi$ with respect to the
node $N$ is $\clTok(\DL(C_N))\times\clTyp(\DL(C_N))$.  Assume that
$\clTok(C_i)$ is finite for each $i$-th child node around a branch and
write $\varphi(N_i)=(\{a_\lambda\}_{\lambda\in\Lambda_i}\vDash\Gamma_i)$.
The integration of $\varphi(N_i)$s around the branch is defined with
values in $\DL(\oplus C_i)$ as follows:
\begin{eqnarray*}
  \mu_O(\langle\varphi(N_i)\rangle_{1\le i\le n})&=&
  (\{a^{(k_\lambda)}_\lambda\}_{\lambda\in\Lambda_O}\vDash\bigvee_{1\le i\le n}
  \overline{\mbox{\rm inc}^i}^\wedge(\Gamma_i)),\\
  \mu_A(\langle\varphi(N_{i^\prime})\rangle_{1\le i^\prime\le n})&=&
  (\{a^{(k_\lambda)}_\lambda\}_{\lambda\in\Lambda_A}\vDash
  \overline{\mbox{\rm conj}}^\wedge(\langle\Gamma_{i^\prime}\rangle_{1\le i^\prime\le n})), \mbox{\rm ~and}\\
  \mu_S(\langle\varphi(N_i)_{1\le i^{\prime\prime}\le n}\rangle)&=&
  (\{a^{(k_\lambda)}_\lambda\}_{\lambda\in\Lambda_S}\vDash
  \overline{\mbox{\rm conj}^\prime}^\wedge(\langle\Gamma_{i^{\prime\prime}}\rangle_{i^{\prime\prime}\in I})),
\end{eqnarray*}
where $\forall i. \Lambda_i\subset\Lambda_O$,
$\forall i^\prime. \Lambda_{i^\prime}\subset\Lambda_A$, and where
$\forall i^{\prime\prime}\in I. \Lambda_{i^{\prime\prime}}\subset\Lambda_S$ for
the index set of the cut sequence of $\varphi(N_i)$s and
$\overline{\mbox{\rm conj}^\prime}$ is a `cut sequence-version' of 
$\overline{\mbox{\rm conj}}$.

\begin{Prop}\label{Integration}
  The function $\varphi$ is a quasi-attribute.  Moreover, the integrated
  effects reflect child nodes' effects.
\end{Prop}

\noindent(Proof) Remark that the index set $\Lambda_i$ of the $i$-th
token $\{a_\lambda\}_{\lambda\in\Lambda_i}$ in $\DL(C_i)$ can be
reduced due to the structural deductions mentioned in
Section~\ref{SectConceptModel}.  Hence we can assume $\Lambda_i$ is
isomorphic to a subset of $\clTok(C_i)$.  In particular,
$\Lambda_i\cap\Lambda_j=\emptyset$ if $i\neq j$.

Around the branch, the values of $\varphi$ for child nodes are integrated
with the use of Proposition~\ref{PropLift} and Proposition~\ref{PropLift2}
as follows.
\begin{itemize}
\item \orb branch: For the $i$-th child's effect
  $(\{a_\lambda\}_{\lambda\in\Lambda_i}\vDash\Gamma_i)$, we observe
  its target $\{a_\lambda\}_{\lambda\in\Lambda_i}$ has a preimage of
  $\overline{\mbox{\rm inc}^i}^\vee$; that is,
  $\{a_\lambda^{(k_\lambda)}\}_{\lambda\in\Lambda}$ where
  $\Lambda_i\subset\Lambda_O$ and both $a_\lambda$ and
  $a^{(k_\lambda)}_\lambda$ point to the same token if
  $\lambda\in\Lambda_i$.  In particular, we can take
  $\{a_\lambda^{(k_\lambda)}\}_{\lambda\in\Lambda_O}$ as the common
  preimage of $\overline{\mbox{\rm inc}^j}^\vee$ for all $j$.
  Therefore, the $i$-th effect
  $(\{a_\lambda\}_{\lambda\in\Lambda_i}\vDash\Gamma_i)$ is lifted to
  $(\{a^{(k_\lambda)}_\lambda\}_{\lambda\in\Lambda_O}\vDash
  \overline{\mbox{\rm inc}^i}^\vee(\Gamma_i))$.  Finally we have
  $(\{a^{(k_\lambda)}_\lambda\}_{\lambda\in\Lambda_O}\vDash\bigvee_{1\le
    i\le n} \overline{\mbox{\rm inc}^i}^\vee(\Gamma_i))$ in
  $\DL(\oplus C_i)$ as the integrated effect.  It is
  $\mu_O(\langle\varphi(N_i)\rangle_{1\le i\le n})$, and obviously, it
  holds if and only if $(\{a^i_\lambda\}_{\Lambda_i}\vDash\Gamma_i)$
  hold for some $i$.
\item \andb branch: Notice that the tuple of child's effects
  $\langle\{a^i_\lambda\}_{\lambda\in\Lambda_i}\vDash\Gamma_i\rangle_{1\le i\le n}$
  is regarded as the relation
  $(\langle\{a^i_\lambda\}_{\lambda\in\Lambda_i}\rangle_{1\le i\le n}\vDash
  \langle\Gamma_i\rangle_{1\le i\le n})$ in $(\DL(C_i))_{1\le i\le n}$.
  We can see that the target
  $\langle\{a^i_\lambda\}_{\lambda\in\Lambda_i}\rangle_{1\le i\le n}$
  has a preimage by $\overline{\mbox{\rm conj}}^\vee$, presented as
  $\{a^{(k_\lambda)}_\lambda\}_{\lambda\in\Lambda_A}$ where $\Lambda_i\subset\Lambda_A$
  for $\forall i$, and where both $a^i_\lambda$ and $a^{(k_\lambda)}_\lambda$
  point to the same token if $\lambda\in\Lambda_i$.  Therefore, the tuple
  $\langle\{a^i_\lambda\}_{\lambda\in\Lambda_i}\vDash\Gamma_i\rangle_{1\le i\le n}$ is
  lifted to $(\{a^{(k_\lambda)}_\lambda\}_{\lambda\in\Lambda_A}\vDash
  \overline{\mbox{\rm conj}}^\wedge(\langle\Gamma_i\rangle_{1\le i\le n}))$
  in $\DL(\oplus C_i)$, and it is
  the integration $\mu_A(\langle\varphi(N_i)\rangle_{1\le i\le n})$.
  As in the previous case, it holds if and only if
  $(\{a^i_\lambda\}_{\lambda\in\Lambda_i}\vDash\Gamma_i)$ hold for all $i$.
\item \sandb branch: Consider the cut sequence of child's effects.
  Similar to the \andb branch, the sequence
  $\langle\{a^r_\lambda\}_{\lambda\in\Lambda_r}\vDash\Gamma_r\rangle_{r\in I}$,
  where $I$ is the index set of the cut sequence, is lifted
  to a relation in $\DL(\oplus C_i)$ and the integration
  $\mu_S(\langle\varphi(N_i)_{1\le i\le n})\rangle$ is defined.
\end{itemize}
Commutativities for quasi-attributes are derived easily.
\EoP

Although we do not consider it in detail in this paper, the
completeness of the decompositions of attacks is defined as follows:
an integrated effect is derived from the parent's effect, that is
$\Gamma^p$ is less than the type part of
$\mu_\sharp(\langle\varphi(N_i)\rangle_{1\le i\le n})$ where
$\sharp\in\{O, A, S\}$.

The following proposition links between the consistency of a branch and
the quasi-attribute of effects.  Owing to Proposition~\ref{Integration},
it is sufficient to consider the relationship between $\varphi(N_p)$
and the integration of child nodes' effects.
As a corollary, we can judge the inconsistency of a branch.

\begin{Prop}\label{PropATValidity}
  Let us consider a consistent branch in an attack tree.
  Suppose the infomorphism realizing the consistency 
  satisfies the following properties:
  the token-part of the infomorphism is determined
  by $\clTok(C_p)$; that is, it preserves the disjoint unions of families.
  Then, there is an infomorphism from an embedded classification with
  the integration of $\langle\varphi(N_i)\rangle_{1\le i\le n}$ to $\DL(C_p)$ such
  that it realizes the original abstraction.
\end{Prop}

\begin{Cor}\label{ATValidity}
  For a branch in an attack tree, if the effect $\varphi(N_p)$ and
  the integration $\mu_\sharp(\langle\varphi(N_i)\rangle_{1\le i\le n})$
  ($\sharp\in\{O,A,S\}$)
  cannot be linked with an infomorphism, then the branch is inconsistent.
\end{Cor}

Rigorously speaking, the corollary negates only infomorphisms with
the properties stated in Proposition~\ref{PropATValidity}.
However, it is sufficient from the viewpoint of refinement.
In our context, the token-part of the infomorphism refines individual objects
in $C_p$ and it is unrealistic that a specific combination
$\{a_1,a_2\}\in\clTok(\DL(C_p))$ is refined to some objects independently
on the refinement of $\{a_1\}$ and $\{a_2\}$.  That is,
$f^\vee(\{a_1,a_2\})=f^\vee(\{a_1\})\sqcup f^\vee(\{a_2\})$ holds.

\paragraph{(Proof of Proposition)}
The required infomorphism is constructed regarding types of branches.

First consider the case of \orb branch.  Suppose the $i$-th effect
$(\{a^i_\lambda\}_{\Lambda_i}\vDash \Gamma_i)$ is a refinement of
$(\{a^p_\lambda\}_{\Lambda_p}\vDash\Gamma_p)$ by an infomorphism
$(f_i^\wedge,f_i^\vee):D_i\rightleftarrows\DL(C_p)$ for each $i$ and
the embedded classification $D_i$.  We define an infomorphism
$(g^\wedge,g^\vee): \DL(\oplus C_i)\rightleftarrows\DL(C_p)$ which
derives the refinement between
$\varphi(N_p)=(\{a^p_\lambda\}_{\Lambda_p}\vDash\Gamma_p)$ and the
integration $\mu_O(\langle\varphi(N_i)\rangle_{1\le i\le n})$.
  
For types, it is enough that $g^\wedge$ is defined for each generator
$(\alpha^{(i)})_\lambda\in\DL(\oplus C_j)$.  This generator is the
embedded type of $\alpha_\lambda\in\clTyp(\DL(C_i))$ by $\overline{\mbox{\rm
    inc}^i}^\wedge$ and we can define
$g^\wedge((\alpha^{(i)})_\lambda)=f_i^\wedge(\alpha_\lambda)$.

For tokens, the image $f_i^\vee(\{c^p_\lambda\}_M)$ for an element
$\{c^p_\lambda\}_M\in\clTok(\DL(C_p))$ equals to
$\bigsqcup_{\lambda\in M}f_i^\vee(\{c^p_\lambda\})$ by assumption, and it
can be written as the form $\{b^i_\lambda\}_{\Lambda_i}\in\clTok(\DL(C_i))$
with a fresh index set $\Lambda_i$.  We can aggregate the images for all $i$
and define $g^\vee(\{c^p_\lambda\}_M)=
\{b^{(k_\lambda)}_\lambda\}_{(\sqcup_j \Lambda_j)}\in\clTok(\DL(\oplus C_j))$ where
$k_\lambda=i$ and $b^{(k_\lambda)}_\lambda=b^i_\lambda$ if $\lambda\in\Lambda_i$.
  
Assume that $g^\vee(\{a^p_\lambda\}_{\Lambda_p})\vDash(\alpha^{(i)})_\mu$ in
$\DL(\oplus C_j)$.  It is rewritten to $\{b^{(k_\lambda)}_\lambda\}_{(\sqcup_j\Lambda_j)}
\vDash(\alpha^{(i)})_\mu$, and is reduced to
$\{b^{(i)}_\lambda\}_{\Lambda_i}\vDash(\alpha^{(i)})_\mu$.  The following equivalence
\begin{eqnarray*}
  \lefteqn{\{b^{(i)}_\lambda\}_{\Lambda_i}\vDash(\alpha^{(i)})_\mu
    ~~~~(\mbox{in $\DL(\oplus C_j)$})}\\
  &\iff&\{b^i_\lambda\}_{\Lambda_i}\vDash\alpha_\mu~~~~(\mbox{in $\DL(C_i)$})\\
  &\iff& f_i^\vee(\{a^p_\lambda\}_{\Lambda_p})\vDash\alpha_\mu~~~~(\mbox{in $\DL(C_i)$})\\
  &\iff& \{a^p_\lambda\}_{\Lambda_p}\vDash f_i^\wedge(\alpha_\mu)~~~~(\mbox{in $\DL(C_p)$})\\
  &\iff& \{a^p_\lambda\}_{\Lambda_p}\vDash g^\wedge(\alpha^{(i)}_\mu)~~~~(\mbox{in $\DL(C_p)$})
\end{eqnarray*}
shows $(g^\wedge,g^\vee)$ is an infomorphism.

The token of the integrated effect $\mu_O(\langle\varphi(N_i)\rangle_{1\le i\le  n})$ is of
the form $\{a^{(k_\lambda)}_\lambda\}_\Lambda$ where $\Lambda_i\subset\Lambda$ and
$a^{(k_\lambda)}_\lambda$ equals to $a^i_\lambda$ in the token part of $\varphi(N_i)$
if $\lambda\in\Lambda_i$.  Namely, this token is
$\bigsqcup_{1\le i\le n} f_i^\vee(\{a^p_\lambda\}_\Lambda)$ which equals to
$g^\vee(\{a^p_\lambda\}_\Lambda)$.

Finally,
\begin{eqnarray*}
  \mu_O(\langle\varphi(N_i)\rangle_{1\le i\le  n})&=&(\{a^{(k_\lambda)}_\lambda\}_\Lambda\vDash
  \bigvee_{1\le i\le n}
  \overline{\mbox{\rm inc}^i}^\wedge(\Gamma_i))\\
  &\iff&g^\vee(\{a^p_\lambda\}_\Lambda)\vDash
  \bigvee_{1\le i\le n}
  \overline{\mbox{\rm inc}^i}^\wedge(\Gamma_i)\\
  &\iff&\{a^p_\lambda\}_\Lambda\vDash g^\wedge(\bigvee_{1\le i\le n}\overline{\mbox{\rm inc}^i}^\wedge(\Gamma_i))\\
  &\iff&\{a^p_\lambda\}_\Lambda\vDash\bigvee_{1\le i\le n} g^\wedge(\overline{\mbox{\rm inc}^i}^\wedge(\Gamma_i))\\
  &\iff&\{a^p_\lambda\}_\Lambda\vDash\bigvee_{1\le i\le n} f_i^\wedge(\Gamma_i)
\end{eqnarray*}
and the type $\bigvee_{1\le i\le n} f_i^\wedge(\Gamma_i)$ in the last is less than
$\varphi(N_p)$.

Next, consider the case of \andb branch (\sandb branch case is proved
in the same way). Suppose the tuple
$\langle\{a^i_\lambda\}_\Lambda\vDash \Gamma_i\rangle_{1\le i\le k}$ is a refinement
of $\{a^p_\lambda\}_{\Lambda_p}\vDash\Gamma_p$ by an infomorphism
$(f^\wedge,f^\vee):(\DL(C_i))_{1\le i\le k}\rightleftarrows\DL(C_p)$.

By Proposition~\ref{PropLift2}, the classification $(\DL(C_i))_{1\le i\le n}$
is embedded into $\DL(\oplus C_i)$, and we denote the embedded image by $D$.
Here,
\begin{eqnarray*}
  \clTok(D)&=&\{\{c^{(k_\lambda)}_\lambda\}_M\ |\ 1\le\forall i\le n.\exists\lambda\in M.k_\lambda=i\}, ~\mbox{and}\\
  \clTyp(D)&=&\{\bigwedge_{1\le i\le n}\Delta^{(i)}_i\ |\ \Delta_i\in\clTyp(\DL(C_i))\}.
\end{eqnarray*}
We can define the infomorphism $(g^\wedge,g^\vee):D\rightleftarrows\DL(C_p)$ by
\begin{eqnarray*}
  g^\wedge(\bigwedge_{1\le i\le n}\Delta^{(i)}_i)&=&f^\wedge(\langle\Delta_i\rangle_{1\le i\le n}),
  ~~\mbox{and}\\
  g^\vee(\{c^p_\lambda\}_M)&=&\bigsqcup_{1\le i\le n}f^\vee(\{c^p_\lambda\}_M)^{(i)}_i.
\end{eqnarray*}
It indeed satisfies the relation (IM) due to the assumption.

Since the family $\{a^i_\lambda\}_{\Lambda_i}$ is the $i$-the component of
$f^\vee(\{a^p_\lambda\}_{\Lambda_p})$, its aggregation
$\bigsqcup_{1\le i\le n}f^\vee(\{a^p_\lambda\}_{\Lambda_p})^{(i)}_i$, the token part of
$\mu_A(\langle\varphi(N_i)\rangle_{1\le i\le n})$, has the preimage by $f^\vee$.
Therefore, $\mu_A(\langle\varphi(N_i)\rangle_{1\le i\le n})$ is lifted to
$\{a^p_\lambda\}_{\Lambda_p}\vDash f^\wedge(\langle\Gamma_i\rangle_{1\le i\le n})$
by (IM) and we can see its type part is less than $\Gamma_p$\EoP

\subsection{Mitigation and Effects}\label{SectMitigation}
Attacks can be treated by countermeasures.  A countermeasures prevents
an attack or modifies its results, and therefore it mitigates the
effect of the attack.  Let us consider the classification $\DL(C)$ for
an attack.  When the original effect is $(\vec{a}\vDash\Gamma)$ and
$\Gamma=\Gamma^\prime\land\Gamma^{\prime\prime}$, we can reduce the
effect to $\vec{a}\vDash\Gamma^\prime$ by canceling the part
$\Gamma^{\prime\prime}$.  This {\it reduction} is also presented as
$\Gamma\le\Gamma^\prime$ with respect to the order of
$\clTyp(\DL(C))$.  For example, if a countermeasure prevents the
effect completely, then the whole of $\Gamma$ is canceled and the
residual part $\Gamma^\prime$ is the top element $\top$.  It can be
interpreted as a valid situation that any primitive type in
$\clTyp(\DL(C))$ does not hold for any token.

Next we consider abstract-refinement relations and the reduction of effects.
Although we can assign a countermeasure to each effect independently,
these countermeasures may break the refinement relation in general.
In order to keep refinement relations, mitigations must be also related.

The next proposition states the maximal mitigation for effects in
the classification of refinement.
Assume $(\vec{a}\vDash_C\Gamma)$ in $\DL(C_C)$ is a refinement of
$(\vec{b}\vDash_A\Delta)$ in $\DL(C_A)$ by $(f^\wedge,f^\vee)$.
\begin{Prop}\label{Mitigate_Refine}
  Consider type $\Delta^\prime$, which is a reduction of $\Delta$.
  If the reduction $\Gamma^\prime$ of $\Gamma$ preserves
  the abstract-refinement relation with $\Delta^\prime$,
  then they are restricted by the inequality
  $f^\wedge(\Gamma^\prime)\lor\Delta\le\Delta^\prime$.
\end{Prop}

The inequality is derived from $f^\wedge(\Gamma^\prime)\le\Delta^\prime$
and $\Delta\le\Delta^\prime$.

By this proposition, type $\Gamma$ cannot be mitigated completely
if $\Delta$ is not as well, for example.

We can observe an application of Proposition~\ref{Mitigate_Refine} to
a branch in an attack tree.  With the same notations used in
Section~\ref{ConstBranch}, the effects of child nodes are written as
$\vec{a}^1\vDash\Gamma_1,\vec{a}^2\vDash\Gamma_2,\dots,
\vec{a}^n\vDash\Gamma_n$, and the effect of the parent node is written
as $\vec{a}^p\vDash\Gamma_p$.  For a consistent branch, if each
$\Gamma_i$ is reduced to $\Gamma_i^\prime$, then the reduced effects
must satisfy the inequality
\[
f^\wedge(\sharp\Gamma_i^\prime)\lor\Gamma_p\le \Gamma_p^\prime
\]
where $\sharp$ is either $\bigwedge$ or $\bigvee$, depending on the
branch type.  Moreover, the LHS is equal to
$\sharp_if^\wedge(\Gamma_i^\prime)$, if $f^\wedge$ preserves $\land$
and $\lor$.  Then, we can state a corollary as follows.

\begin{Cor}\label{Mitigate_Refine_AT}
  For a consistent \orb branch, if the residual effects preserve
  consistency, then none of $f^\wedge(\Gamma^\prime_i)$s is greater than
  $\Gamma^\prime_p$.  In other words, the mitigated effect of the parent node
  is weaker than that of each child node.
\end{Cor}

However, the same cannot be said for \andb branches.
There may exist $\Gamma_i^\prime$s such that each $f^\wedge(\Gamma_i^\prime)$ is not
less than $\Gamma_p^\prime$ while their conjunction
$\bigwedge_if^\wedge(\Gamma_i^\prime)$ is less than $\Gamma_p^\prime$.
Furthermore, another factor prevents us from consideration for \sandb branches.
The effects of attacks in a \sandb branch depend on the preceding effects and
attacks, and thus mitigations for effects are constrained by 
the relationship with not only a parent's effect but also the preceding ones.

\section{Case Study}\label{CStudy}
Here, we apply the ideas described in Section~\ref{EffTr} to practical
situations.  The first case study is about improving an attack tree
(Section~\ref{EgStr}).  We make explicit invalid branches in an attack
tree with effects and propose correct decompositions for them.  The
second case study concerns mitigating effects
(Section~\ref{EvalSubSect}).  Consistency is given for an attack tree
rigorously, and possible degrees of mitigation are considered.
Beforehand, the entire security analysis and the target system are
described as a context.

\subsection{Process Overview}\label{TPoverviewSubSect}
The security analysis process specified in JASO TP15002~\cite{TP15002}
consists of five phases: ToE (Target of Evaluation) definition, Threat
analysis, Risk assessment, Define security objectives, and Security
requirement selection.  We follow a concretization of it studied
in~\cite{SCN19}.  It also reflects a refactoring of the process and
data in the original security analysis process.

A deliverable of Risk assessment phase is a list of identified threats
to ToE.  Threats in this list are prioritized, and significant ones
are analyzed in depth using tree diagrams.  Then, countermeasure goals
for the threats are discussed in Define security objectives phase.

\subsection{Target of Evaluation(ToE)}\label{ToESubSect}
\begin{figure}
  \begin{center}
    \includegraphics[keepaspectratio,width=100mm]{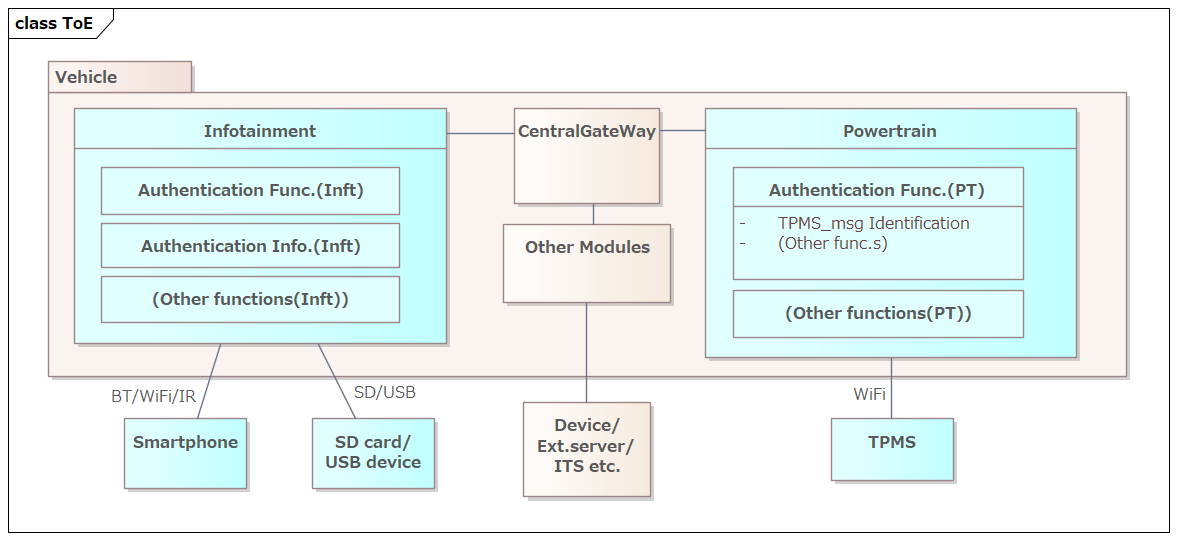}
  \end{center}
  \caption{The network architecture of ToE}\label{ToE}
\end{figure}

Based on the model in~\cite{MYKMHK}, a vehicular network system can be
specified as the ToE.  While the entire network system is analyzed
in~\cite{SCN19}, the focus of this paper is on specific parts of the
network (Fig.~\ref{ToE}).  Concretely, in Section~\ref{EgStr}, we
consider the identified threat ``{\it Authentication function in
  Powertrain is interfered via a Tire Pressure Monitoring
  System(TPMS)}.''  Similarly, in Section~\ref{EvalSubSect}, we
consider ``{\it Authentication information in Infotainment is stolen
  via smartphone}.''  Both threats are estimated to be significant,
since the target modules are close to their entry points and the
potential damages to the vehicle system are severe.

\subsection{Verifying Decompositions}\label{EgStr}
\begin{figure}[bpht]
  \begin{center}
    \parbox{100mm}{\scriptsize
      {\bf Node labels}:\\
      ([AuF.PT], [msgId.PT] mean `Authentication function in Powertrain,'
      `TPMS message identification function in Powertrain'.)\\
      A0: [AuF.PT] is interfered with via TPMS.\\
      A1: [msgId.PT] is tampered with.\\
      A1.1: Powertrain Software is tampered with.\\
      A2: [msgId.PT] is interfered with by DoS.\\
      \vspace{\parsep}
      
      {\bf Effects}\\
      E0: $\symbolw{AuthF\_PT}\vDash\symbolw{UI\_behavior}$
      (Unintended behaviors in Authentication function of Powertrain)\\
      E1: $\symbolw{MsgIdF\_PT}\vDash\symbolw{invalid}$
      (Installed TPMS msg identification function is invalid)\\
      E1.1: $\symbolw{Software\_PT}\vDash\symbolw{invalid}$
      (Installed Powertrain Software is invalid)\\
      E2: $\symbolw{MsgIdF\_PT}\vDash\symbolw{unavailable}$
      (TPMS msg identification function is not available)
    }\\
    \begin{tabular}{cc}
      \includegraphics[keepaspectratio,height=28mm]{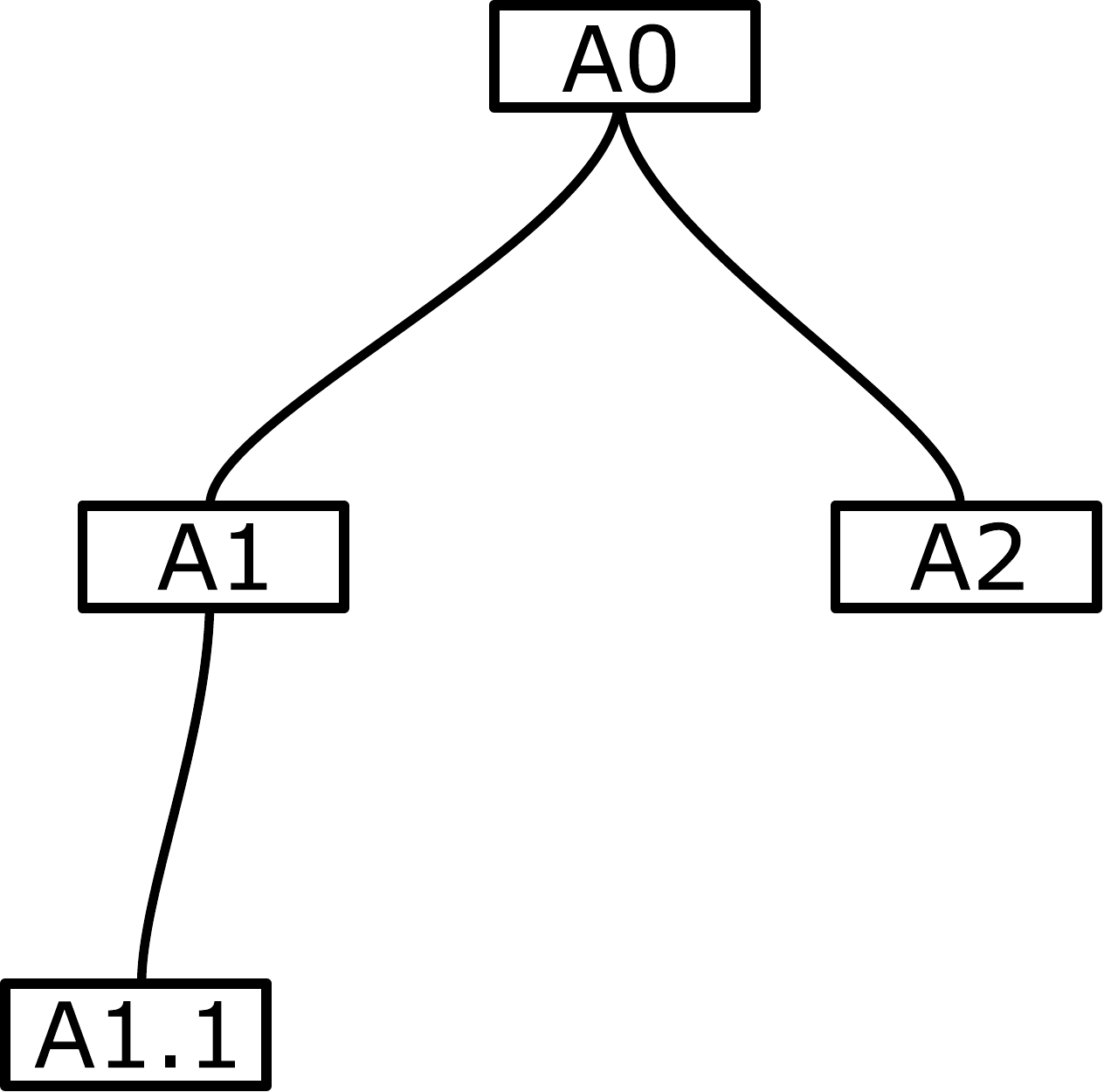}~&~
      \includegraphics[keepaspectratio,height=28mm]{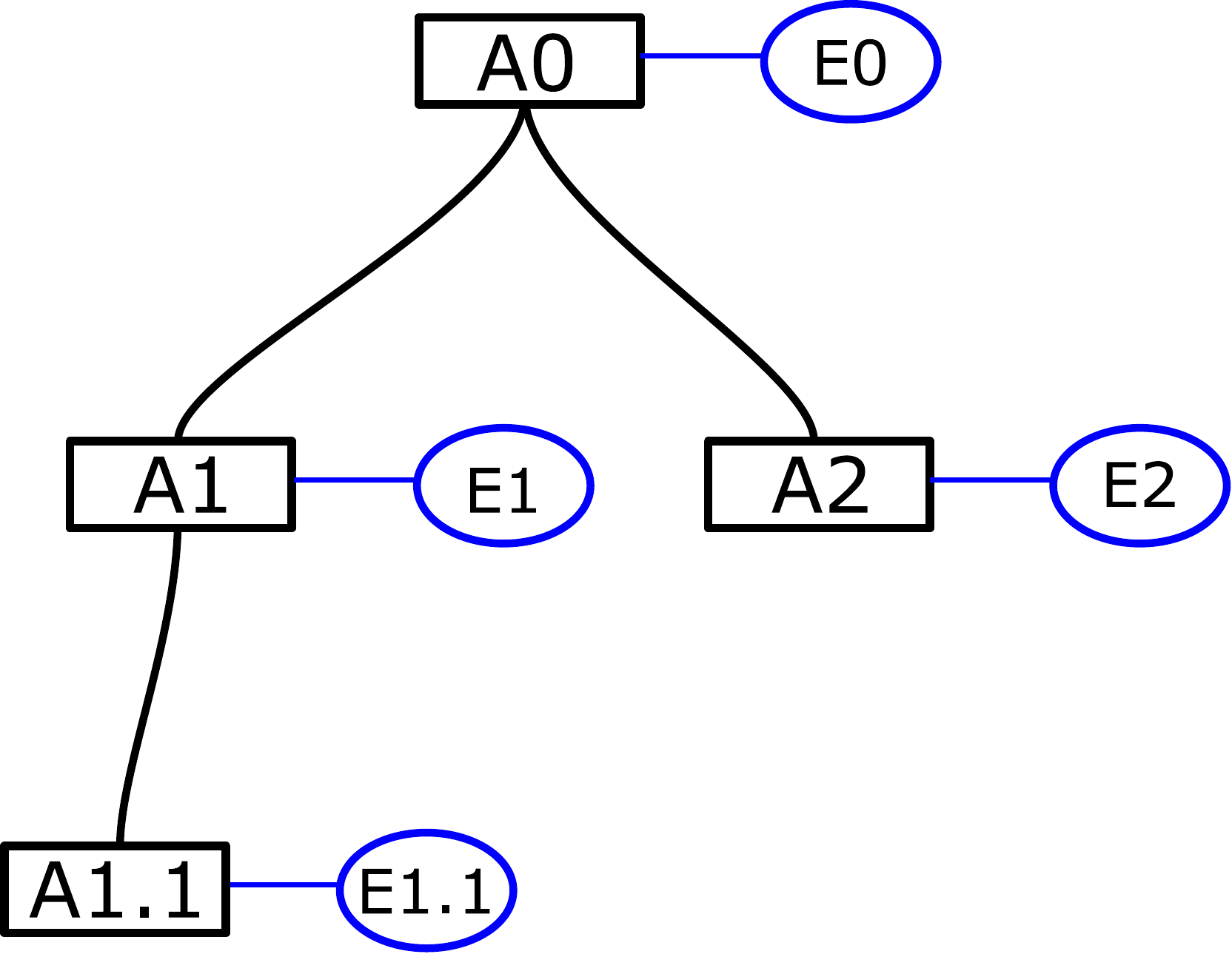}\\
      (a)Early version ~&~
      (b)Early version with effects
    \end{tabular}\\
    \vspace*{3mm}
    \parbox{100mm}{\scriptsize {\bf Revised node labels}:\\ A1:
      Unauthorized [msgId.PT] is invoked.\\
      A1.1: [msgId.PT] is tampered with.\\
      A1.1.1: [msgId.PT] is tampered with by replacing Powertrain
      software.\\
      A1.2: (Unauthorized) [msgId.PT] is invoked.\\
      \vspace{\parsep}
      
      {\bf Revised effects}:\\
      E1: $\symbolw{MsgIdF\_PT}\vDash\symbolw{UI\_behavior}$
      (Unintended behaviors in TPMS msg identification function)\\
      E1.1: $\symbolw{MsgIdF\_PT}\vDash\symbolw{invalid}$
      (Installed Powertrain Software is invalid)\\
      E1.1.1: $\symbolw{MsgIdF\_PT}\vDash\symbolw{invalid}$
      (Installed TPMS msg identification function is invalid)\\
      E1.2: $\symbolw{MsgIdF\_PT}\vDash\symbolw{UI\_behavior}$
      (Unintended behaviors in TPMS msg identification function.
    }\\
    \includegraphics[keepaspectratio,height=38mm]{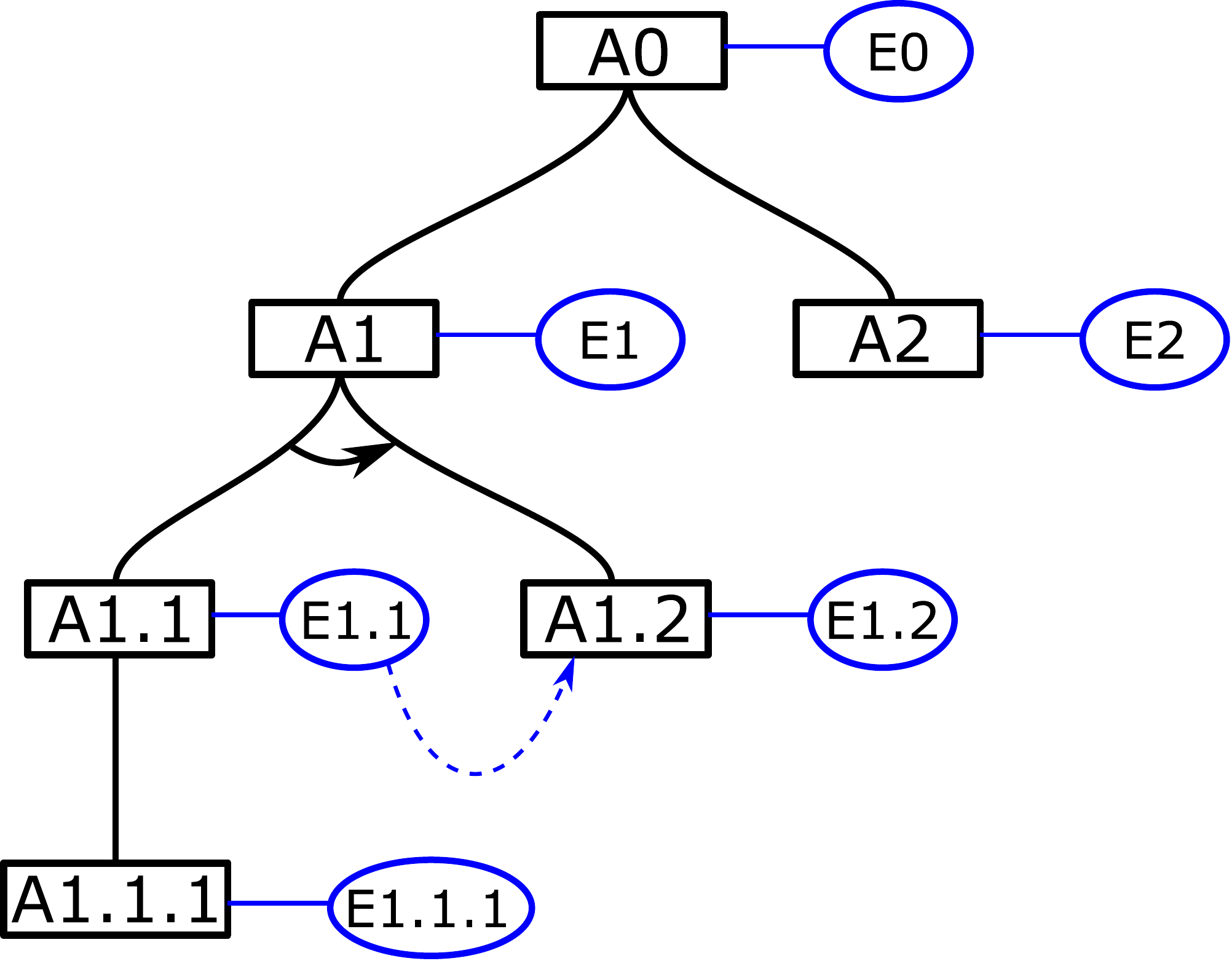}\\
    (c)Improved attack tree
  \end{center}
  \caption{Improvement of an attack tree}\label{AT7_1}
\end{figure}

In this subsection, we consider the threat ``{\it Authentication
  function in Powertrain is interfered via TPMS.}''
Fig.~\ref{AT7_1}(a) shows an early version of the attack tree for the
threat (only the upper part is outlined).  It should be noted that
attacks on node labels include the events that contribute to invoking
the parent but are not performed by the attacker.  Policies and
methods used to construct the tree are slightly ambiguous, and the
structures of modules (Fig~\ref{ToE}) are not considered well.
Therefore, sub-attacks of an attack are intuitively selected; some of
the sub-attacks do not refine the parent but are expected to occur
preceding the parent.  Here, decomposition is interpreted as causal
ones implicitly.  Moreover only \orb and \andb branches are
considered.  As a result, the following two inconsistencies are found:
\begin{enumerate}
\item A temporal gap among attacks around a branch.  Attack A1 has
  to occur before its parent A0, but attack A2 and its parent A0
  occur simultaneously.
\item A violated refinement order.  The Msg. identification function
  in A1 refines the Authentication function in A0, whereas Powertrain
  Software mentioned in A1.1 is a wider entity than
  Msg. identification function in A1.
\end{enumerate}

These un-structural situations are made explicit by considering the
effects of attacks.  As mentioned in Section~\ref{EffTr}, effects are
changes in ToE and its environment caused by specific actions.  Here,
we consider the entities affected by attacks for all nodes and
identify the effects on them derived from the attacks.  The results
are illustrated in Fig~\ref{AT7_1} (b).

First, let us examine the relations between A0 and its children A1 and
A2.  The correspondences of tokens are appropriate; namely, TPMS msg
identification function presented by \symbolw{MsgIdF\_PT} is a refined
entity of Authentication function.  On the other hand, we observe that
an unintended behavior expresses a dynamic aspect of the function
while invalidness is a static one.  The invalidness cannot be related
to the unintended behavior semantically.  Therefore, we conclude that
no abstract-refinement relation exists between E0 and E1 and the
branch around A0 is inconsistent, if we consider infomorphisms
reflecting ToE.

Next, let us move on to the branch around A1.  Clearly Powertrain Software
is a higher concept in Fig.~\ref{ToE} and we cannot define any mapping
of tokens that maps \symbolw{MsgIdF\_PT} to \symbolw{Software\_PT}.
The branch is inconsistent as we cannot construct infomorphisms realizing
the abstract-refinement relation.

The revised version of the attack tree is given in
Fig.~\ref{AT7_1}(c).  The second inconsistency mentioned above is
resolved by emphasizing the target (A1.1.1).  We modify node A1 and
its subtree to ensure consistency.  The targets of nodes in the
subtree are unified to TPMS msg identification function presented by
\symbolw{MsgIdF\_PT}, yielding the common classification for
considering their effects; in particular, token parts of infomorphisms
realizing abstract-refinement relations are identity maps.  Moreover,
the cut sequence of effects E1.1 and E1.2 consists of a single
relation $\symbolw{MsgIdF\_PT}\vDash \symbolw{UI\_behavior}$.  At
last, all branches are consistent in the tree, and the entire attack
tree is consistent as well.

\subsection{Degrees of Possible Mitigation}\label{EvalSubSect}
\begin{figure}
  \begin{center}
    \includegraphics[keepaspectratio,width=60mm]{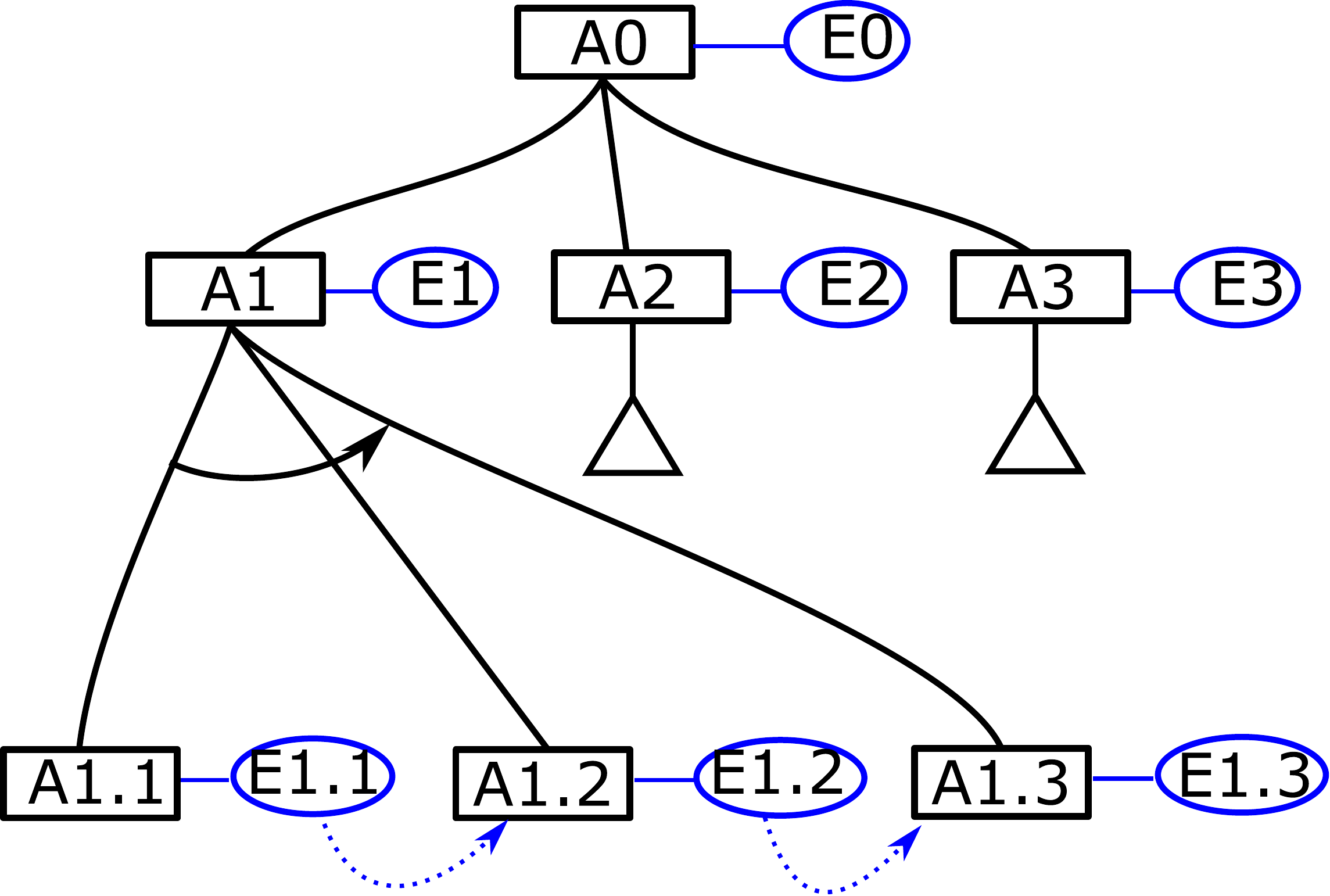}\\~\\
    \parbox{100mm}{\scriptsize
      {\bf Trianglar Nodes:}\\
      The triangles connected to A2 and A3 indicates their
      decompositions are left undeveloped.\\
      {\bf Node Labels:}\\
      ([AuI.I] and [AuF.I] means Authentication Information and
      Authentication Function in Infotainment respectively.)\\
      A0: [AuI.I] is stolen via BT/Wifi/IR (smartphone).\\
      A1: [AuI.I] is stolen by reverse engineering.\\
      A2: [AuI.I] is obtained by brute-force.\\
      A3: [AuI.I] is obtained by eavesdropping BT/Wifi/IR.\\
      A1.1: Procuring a device which had connected to the target.\\
      A1.2: Analyzing the device.\\
      A1.3: Identifying [AuI.I].\\
      {\bf Effects:}\\
      E0, E1, E2, E3, E1.3: $\symbolw{AuI.I}\vDash\symbolw{Disc}$ ([AuI.I] is disclosed)\\
      E1.1: $\symbolw{Data}\vDash\symbolw{Acc}$
      (Data in the device is accessible)\\
      E1.2: $\symbolw{Data}\vDash\symbolw{Disc}$ (Data in the device is
      disclosed)
    }
  \end{center}
  \caption{Attack tree for ``Authentication information is stolen''
    }\label{ATwET}
\end{figure}

In this subsection, we consider the threat ``{\it Authentication information in
  Infotainment is stolen via BT/Wifi/IR (smartphone).}''  The attack tree
for the threat is depicted in Fig.~\ref{ATwET}.  In particular, we focus on
branch A1, where reverse engineering is attempted on the mobile device that has
connected to the Infotainment in the past.

\begin{table}
\caption{Token and types to describe effects for the tree in
  Fig.~\ref{ATwET}}\label{ClaElem}
  \begin{center}{\footnotesize
  \begin{tabular}{|l|l|}
    \multicolumn{2}{c}{The classification $C_0$ [re. $C_1$, $C_{1.3}$]
      for Node A0 [re. A1, A1.3]}\\\hline
    Token & Type\\\hline
    \symbolw{AuI.I}, \symbolw{AuF.I},&\symbolw{Disc} (Disclosed),
    \symbolw{Acc} (Accessible),\symbolw{Mod} (Modified),\\
     ($\varepsilon$)&\symbolw{Inv} (Invalid),\symbolw{Unav} (Unavailable),
    \symbolw{Ubhv} (Unintended behavior)\\\hline
    \multicolumn{2}{|c|}{Token-Type relations}\\\hline
    \multicolumn{2}{|c|}{$x\vDash\symbolw{Disc}$, $x\vDash\symbolw{Acc}$~~
    ($x\in\{\symbolw{AuF},\symbolw{AuI}\}$)}\\\hline
  \end{tabular}
  
  \vspace*{5mm}
  \begin{tabular}{|l|l|}
    \multicolumn{2}{c}{The classification $C_{1.1}$ [re. $C_{1.2}$]
      for Node A1.1 [re. A1.2]}\\\hline
    Token & Type\\\hline
    \symbolw{Mech} (mechanical part),&\symbolw{Disc}, \symbolw{Acc},
    \symbolw{Mod},\\
    \symbolw{Data}, \symbolw{Pgm}, ($\varepsilon$)&\symbolw{Inv},\symbolw{Unav},
    \symbolw{Ubhv}\\\hline
    \multicolumn{2}{|c|}{Token-Type relations}\\\hline
    \multicolumn{2}{|c|}{$x\vDash\symbolw{Disc}$, $x\vDash\symbolw{Acc}$,
    $\symbolw{Mech}\vDash\symbolw{Acc}$~~
    ($x\in\{\symbolw{Data},\symbolw{Pgm}\}$)}\\\hline
  \end{tabular}}
  \end{center}
\end{table}

Primitive tokens and types to describe effects are listed in
Table~\ref{ClaElem}.  Just two kinds of classifications are shown,
since plural nodes refer their tokens and types by the same names.  We
assume the order $\symbolw{Disc}<\symbolw{Acc}$, meaning that
Information disclosure includes the situation that the information is
accessible (but not vice versa), while orders of other pairs of
primitive elements are not cared.

First, we check that the branch at node A1 is consistent by defining
an infomorphism $(f^\wedge,f^\vee)$.  Attack A1.2 (analyzing the
device) can be done under effect E1.1 (data in the device are
accessible), and attack A1.3 can be done under effect E1.2 as well.
Since the cut sequence of the effects of child nodes A1.1, A1.2 and
A1.3 is $(\symbolw{Data}\vDash\symbolw{Disc},
\symbolw{AuI.I}\vDash\symbolw{Disc})$, the infomorphism is
$(\DL(C_{1.2}),\DL(C_{1.3}))\rightleftarrows\DL(C_1)$, and we can
define the token part and the type part as follows.
\[
f^\wedge(\langle x,y\rangle)=\left\{\begin{array}{ll} \symbolw{Disc}
&~ (\langle
x,y\rangle=\langle\symbolw{Disc},\symbolw{Disc}\rangle)\\ \symbolw{Acc}
&~ (\langle x,y\rangle=\langle\symbolw{Disc},\symbolw{Acc}\rangle,
\langle\symbolw{Acc},\symbolw{Disc}\rangle,\mbox{~or~}\langle\symbolw{Acc},
\symbolw{Acc}\rangle)\\ \top
&~ (\mbox{otherwise})
\end{array}\right.
\]
and the mapping is extended for compound types naturally.  For tokens,
all except one token are mapped to the trivial pair.
\[
f^\vee(x)=\left\{\begin{array}{ll}
\langle\{\symbolw{Data}\},\{\symbolw{AuI.I}\}\rangle&~(x=\{\symbolw{AuI.I}\})\\
\langle\{\varepsilon\},\{\varepsilon\}\rangle &~ (\mbox{otherwise})
\end{array}\right.
\]
By $(f^\wedge,f^\vee)$, the cut sequence mentioned above is equivalent to
$(\symbolw{AuI.I}\vDash\symbolw{Disc})$ in $\DL(C_1)$, which is the
effect of node A1.  Therefore, $(f^\wedge,f^\vee)$ realizes refinement
and the branch is consistent.

Now let us consider reducing effect E1.2 to
$(\symbolw{AuI.I}\vDash\symbolw{Acc})$.
The definition of $f^\wedge$ indicates that the type parts of E1.2 or E1.3 are
reduced to \symbolw{Acc} in order to keep the abstract-refinement relation.
Moreover, the dependency between them restricts the situation that E1.2 remains
unchanged.  Hence, the possible mitigated effects are that the type of
E1.3 is reduced to \symbolw{Acc}.  Finally, as a mitigation, we can encrypt
authentication information in the device so that it is not identified.

\section{Concluding Remarks}\label{Conclusion}
Attack trees are a major tool for security analysis, as they allow
us to formally derive qualitative/quantitative properties of attacks.
However, the attack decompositions in attack trees have been
justified only by reviews.  The chains of formalism in attack tree
processes are unlinked at the construction of the trees.

At first in this paper, we defined a new formal system of attack trees
(with sequential conjunctions).  It is a generalization of the existing
framework~\cite{SeqConjHMT}, and an attack tree is interpreted as
a set of its refinement scenarios.

Next, we proposed a validation of decompositions in attack trees.  An
effect is considered for every attack, and based on those effects, the
consistency of a branch is defined.  The framework for describing
effects applies the theory of the classifications with infomorphisms.
We sought to verify the abstract-refinement relations around branches
in attack trees rigorously and to analyze the structure of the attack.
The chains of formalism are linked with each other, resulting in a
complete application of the formal approach to security analysis.

One application of the framework relates to mitigation of attacks.
A mitigation for an attack cancels some part of the original effect of the
attack.  Possible degrees of mitigation can be measured with effects.

Several issues remain for future research:
\begin{itemize}
\item From a practical viewpoint, the simplification of attack trees is
  important.  Equivalence relations between trees can give us a solution.
  Indeed, the frameworks~\cite{MO, SeqConjHMT} introduced the equivalences
  and discussed detailed properties with transforming attack trees.
  Though the formal system of attack trees in this paper is relatively rigid,
  appropriate equivalence and transformation will improve security analysis
  with attack trees.
\item The discussion on mitigation can be improved.  In this paper, we
  considered the distributive lattice structures of $\clTyp(\DL(C))$.
  The description is weak in the sense that the canceled part of the
  original effect is not always identified.  Namely, not all
  $\Gamma\in\clTyp(\DL(C))$ can be expressed in the form
  $\Gamma^\prime\land \Gamma^{\prime\prime}$.  Hence, we can operate
  only the residual part.  When we focus on the specified effects
  occurring in a target system, we may identify the canceled part like
  the above and evaluate the possible degrees of mitigation more
  concretely.
\item We can verify the consistency of a branch in an attack tree with
  effects, but how do we decompose an attack accurately?  In the case
  study, we show a way to do it following the structure of the target
  system.  It seems to work well in many cases but is not
  comprehensive.  In particular, it is difficult to find sequential
  attacks (for \sandb branches) by observing the static structure.
  Methodologies deriving attack decompositions systematically are
  expected.
\item Tool support.  Defining and managing classifications for every node
  in an attack tree with infomorphisms is a tedious work.  Software for
  attack trees with effects will improve the quality of security analyses.
\end{itemize}

\appendix
\section{Projection on Causal Attack Trees}\label{SectProjCAT}
\subsection{Intermediate Semantics of Attack Trees}\label{HMTDef}
Here, we review the formal definition of attack trees proposed in
Section 4 of~\cite{SeqConjHMT}.  Attack trees are binary trees with
three types of branches, and they allow several equivalent
transformations.  They are referred to as {\it causal attack trees},
and we also use this term in order to distinguish them from our
definition.  Causal attack trees are interpreted as sets of directed
graphs such that their vertices are labeled with the attacks of leaf
nodes in the original trees.

\begin{Def}
  A causal attack tree is a binary tree constructed from the following
  rule:
  \[
  t ::= a~|~ t \triangle t ~|~ t\triangledown t ~|~ t\cdot t,
  \]
  where $a\in\mbox{Atom}$, the set of symbols presenting atomic attacks.
\end{Def}
The operator $\triangle$ [re. $\triangledown$, $\cdot$] corresponds
to an \andb [re. \orb, \sandb] branch.  The associativity and
commutativity for $\triangle$ and $\triangledown$ are assumed, and
distributivity $(t\triangledown u)\triangle v=(t\triangle v)
\triangledown(u\triangle v)$ and idempotency $u\triangle u=u$ as well.
However, for $\cdot$, only associativity and distributivity with \orb,
i.e., $t\cdot(u\triangledown v)=(t\cdot u)\triangledown(t\cdot v)$ and
$(u\triangledown v)\cdot t=(u\cdot t)\triangledown(v\cdot t)$, are
assumed.

\begin{figure}[bpht]
  \begin{center}
    \includegraphics[keepaspectratio,width=100mm]{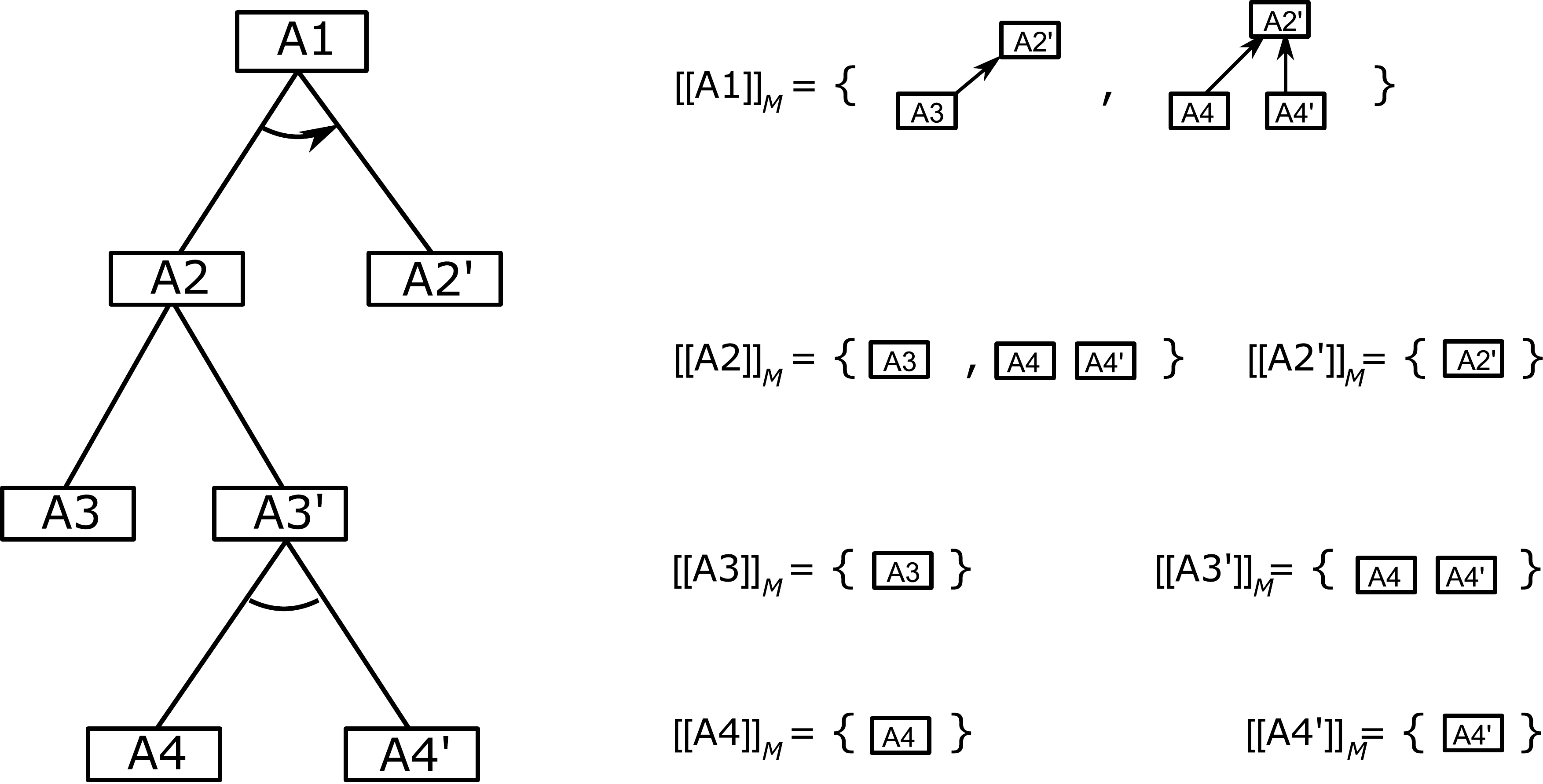}
    \caption{Computation of the semantics of a causal attack tree}\label{EgCausal}
  \end{center}
\end{figure}

Although two kinds of semantics are introduced to deal with the
specialization of the tree effectively, they are induced from the
following generic semantics.
\begin{Def}
  The intermediate semantics of causal attack trees $\semParM{t}$,
  taking values in the set of directed graphs whose vertices are
  labeled with \mbox{Atom}, is defined as follows:
  \begin{itemize}
  \item $\semParM{a} = \{G_a\}$, where $G_a$ is the graph consisting
    of a single vertex labeled with $a$ and no edge.
  \item $\semParM{t_1\triangledown t_2}=\semParM{t_1}\cup\semParM{t_2}$.
  \item $\semParM{t_1\triangle t_2}=\{\tau_1\sqcup\tau_2 ~|~
    \tau_1\in\semParM{t_1}, \tau_2\in\semParM{t_2}\}$,
    where the juxtaposition of graphs $S$ and $T$ is denoted by $S\sqcup T$.
  \item $\semParM{t_1\cdot t_2}$ is the set of pointwise sequential
    composition of $\semParM{t_1}$ and $\semParM{t_2}$.
    i.e. the set of graphs constructed with every pair of
    $g_1\in\semParM{t_1}$ and $g_2\in\semParM{t_2}$,
    where these two graphs are juxtaposed, and all of possible edges from
    $g_1$'s vertices to $g_2$'s vertices are added.
  \end{itemize}
\end{Def}
Fig.~\ref{EgCausal} illustrates an interpretation of a causal attack tree.

\subsection{The Projection to Causal Attack Trees}\label{ProofProjCAT}
Here, we see the relationship between attack trees in \atSet and
causal attack trees defined in Section~\ref{HMTDef}.

First, R-trees, the atomic refinement scenarios of attack trees, are
projected to directed graphs.
\begin{Def}
  The projection $\pi$ from $\atSet_R$ to the set of directed
  graphs with labels is defined as follows:
  \begin{itemize}
  \item $\pi(Lf(a))=G_a$.  I.e. the R-tree having only the root node
    is projected to the singleton graph.
  \item $\pi(Nd(n,\andb,\langle \tau_1,\dots,\tau_m\rangle))
    =\pi(\tau_1)\sqcup \ldots\sqcup\pi(\tau_m)$.
  \item $\pi(Nd(n,\sandb,\langle \tau_1,\dots,\tau_m\rangle))$ is the
    graph $\pi(\tau_1)\sqcup \ldots\sqcup\pi(\tau_m)$ with additional edges
    $\{(v^{(i)},w^{(i+1)})~|~v^{(i)}\in\pi(\tau_i),~ w^{(i+1)}\in\pi(\tau_{i+1}),~
    1\le i\le m-1\}$.
  \end{itemize}
  We abuse the symbol $\pi$ as a function on the set of attack trees.
  Namely $\pi(\{t_1,t_2,\dots,t_m\})=\{\pi(t_1),\pi(t_2),\dots,\pi(t_m)\}$.
\end{Def}

Attack trees in \atSet can be interpreted with intermediate semantics
for causal attack trees.  Notice that it is easy to transform an
attack tree into a causal one.  Indeed, a sequence of each node's
children can be expressed as nested binary terms (dummy labels are
required for intermediate nodes).  This transformation is denoted by
$\beta$.  By the following proposition, semantics are linked as well.

\begin{Prop}\label{PropProjCAT}
  The following diagram commutes:
  \[\begin{CD}
  \atSet @>\beta>> \{\mbox{causal attack trees}\}\\
    @V{\semant{\cdot}}VV @VV{\semParM{\cdot}}V\\
    \mbox{Pow}(\atSet_R) @>{\pi}>> \mbox{Pow}(\{\mbox{directed graphs with labels}\})
  \end{CD}\]
  where $\mbox{Pow}(X)$ means the power set of $X$.
\end{Prop}

\paragraph{Proof.}
We can check the proposition for the type of the top branch.
Denote a sequence $\langle t_1,\dots,t_m\rangle$ by $\bar t$.

Let us take the element $Lf(a)$, the tree having only the root node.
The equalities $\pi(\semant{Lf(n)})=\pi(\{Lf(n)\})=\{G_n\}$ and
$\semParM{\beta(Lf(n))}=\semParM{Lf(n)}=\{G_n\}$ indicates
the proposition.

For the compound attack tree, the proof is divided with respect to
the type of the branch:
\begin{itemize}
\item For an \andb branch, 
  \begin{eqnarray*}
  \pi(\semant{Nd(n,\andb,\bar{t})})&=&
  \pi(\{Nd(n,\andb,\langle \tau_1,\dots,\tau_m\rangle)~|~
  (\tau_1,\dots,\tau_m)\in\semant{\,\bar{t}\,}\})\\
  &=& \{\tau_1\sqcup\dots\sqcup\tau_m~|~
  (\tau_1,\dots,\tau_m)\in\semant{\,\bar{t}\,}\}
  \end{eqnarray*}
  holds, and each element in RHS appears in $\semParM{\beta(Nd(n,\andb,\bar{t}))}$.
\item For a \sandb branch, the proposition is proved in similar way.
  An element in $\pi(\semant{Nd(n,\sandb,\bar{t})})$ is of the form
  $\tau_1\sqcup\dots\sqcup\tau_m$ with additional edges, and it also
  appears in $\semParM{\beta(Nd(n,\sandb,\bar{t}))}$.
\item For an \orb branch,
  \begin{eqnarray*}
    \pi(\semant{Nd(n,\orb,\bar{t})})&=&
    \pi(\bigsqcup_{1\le i\le m}\{Nd(n,\andb,\langle\tau\rangle)|\tau\in\semant{t_i}\})\\
    &=&\bigsqcup_{1\le i\le m}\{\pi(Nd(n,\andb,\langle\tau\rangle))|\tau\in\semant{t_i}\})\\
    &=&\bigsqcup_{1\le i\le m}\pi(\semant{t_i})
  \end{eqnarray*}
  holds.  It equals to $\semParM{\beta(Nd(n,\orb,\bar{t}))}$.\EoP
\end{itemize}

\section*{Acknowledgments}
The authors express our gratitude to Yuichiro Hosokawa, who
let us know about Information flow and Channel theory, and Kenji Taguchi,
who gave us
fruitful comments in the very early phase of this work.
The authors are also grateful to members of SEI-AIST Cybersecurity
Cooperative Research Laboratory for their helpful discussion and kindness.
In particular, Yoichi Hata and Akira Mori gave us continuous encouragement.

\bibliographystyle{fundam}
\bibliography{FundamInfo}

\end{document}